\documentclass[journal]{IEEEtranTCOM}

\normalsize

\usepackage{enumerate}
\usepackage{amssymb}
\usepackage{amsmath}
\usepackage{amsthm}
\usepackage{booktabs}
\usepackage{caption}
\usepackage{subcaption}
\usepackage{float}
\usepackage{latexsym}
\usepackage{algorithm}
\usepackage[noend]{algpseudocode}
\usepackage{algorithmicx}
\usepackage{multirow}
\usepackage{array}
\usepackage{alphalph}
\usepackage{epsfig}
\usepackage{setspace}
\usepackage{cite}
\usepackage{cleveref}
\usepackage{graphicx}
\usepackage{epstopdf}
\usepackage{color,soul}
\usepackage{url}

\renewcommand\IEEEkeywordsname{Index Terms}

\begin{document}
\title{Uplink resource allocation optimization for user-centric cell-free MIMO networks}

\author{Zehua~Li,~\IEEEmembership{Student~Member,~IEEE,}
        and~Raviraj~Adve,~\IEEEmembership{Fellow,~IEEE}
\thanks{The authors are with the Department of Electrical and Computer Engineering, University of Toronto, Ontario, ON M5S 3G4, Canada. E-mails: samzehuali.li@mail.utoronto.ca, rsadve@ece.utoronto.ca.}}

\markboth{IEEE Transactions on Wireless Communications}%
{Submitted paper}
\maketitle

\begin{abstract}

We examine the problem of optimizing resource allocation in the uplink for a user-centric, cell-free, multi-input multi-output network. We start by modeling and developing resource allocation algorithms for two standard network operation modes. The centralized mode provides high data rates but suffers multiple issues, including scalability. On the other hand, the distributed mode has the opposite problem: relatively low rates, but is scalable. To address these challenges, we combine the strength of the two standard modes, creating a new semi-distributed operation mode. To avoid the need for information exchange between access points, we introduce a new quality of service metric to decentralize the resource allocation algorithms. Our results show that we can eliminate the need for information exchange with a relatively small penalty on data rates.

\end{abstract}


\IEEEpeerreviewmaketitle

\section{Introduction} \label{chapter:intro}

With rapid growth in the wireless connectivity market, service providers are constantly seeking ways to deliver higher data rates to a denser population of users. In traditional cellular networks, one of the main factors limiting data rates is interference from neighboring transmitters. Cell-free networks, comprising distributed access points (APs) have been proposed as an alternative; in the popular user-centric version, each user is served by its closest APs~\cite{Demir2021}. Here, we concern ourselves with resource allocation in the uplink of a cell-free user-centric multiple-input, multiple-output (MIMO) network. Resource allocation in the uplink is significantly different from the downlink since the interference pattern changes drastically with scheduling and allocation decisions. 

\subsection{Literature Review}
Resource allocation optimizations are crucial to maximize network performance. It involves maximizing a utility under operating constraints; one of the most popular utility to maximize is the spectral efficiency (SE) or sum-rate across users. An popular variant is the weighted sum SE or weighted sum-rate (WSR) because it not only considers the overall network performance but also takes into account of fairness. In~\cite{khan2020} the authors studied techniques to jointly optimize the scheduling of users and beamformers in the downlink of a cellular network and proposed a  signal-to-inteference-plus-noise ratio (SINR)-based resource allocation algorithm which uses fractional programming (FP) and Hungarian algorithm to maximize the WSR. Since interference is coupled to the decisions made across all base stations (BSs), SINR-based approaches require the BSs to know the channel state information (CSI) of the entire network; this implies an impractically large overhead and is, therefore, not scalable. 

Researchers have proposed alternative approaches based on the signal-to-leakage-plus-noise ratio (SLNR) metric~\cite{liu2014} where power leaked to other cells is treated as a proxy for interference. SLNR-based algorithms allow for decentralized implementation since, at each BS, the SLNR only depends on local decisions. However, optimizing SLNR has important drawbacks such as ineffective power allocations and difficult local scheduling. To combine the strengths of the two metrics, in~\cite{li2022} and~\cite{gamvrelis2022} we proposed a new metric called signal-to-leakage-plus-interference-plus-noise ratio (SLINR), combining intra-cell interference and inter-cell leakage; our SLINR-based approach provides a \textit{decentralized} solution to effectively optimize resource allocation. 


While previous work studied resource allocation of the cellular network in the downlink, the authors of~\cite{shen2018} analyzed the problem in the uplink. Both~\cite{khan2020} and~\cite{shen2018} optimized the user scheduling and beamforming by maximizing for the WSR problem. Common techniques are used such as block coordinate descent, quadratic transform, and Hungarian algorithm. However, the uplink transmit-receive relation is very different from downlink; crucially, in the uplink the interference pattern changes as the scheduling changes. As a result, to maximize the WSR, in~\cite{shen2018} the authors develop a utility value to decouple the relation between scheduling and the data rate and use it as the criteria in the Hungarian algorithm. 

In recent years, the focus is starting to shift from the cellular paradigm to a cell-free network architecture. The concept of user-centric cell-free networks has gathered significant attention~\cite{ammar2022ucsurvey}. In the user-centric setup, each user is served by a cooperative cluster of spatially distributed antennas. Effectively, users are surrounded by antennas as opposed to the conventional scenario where the BSs are surrounded by users. This framework eliminates cell edges and provides uniform coverage and performance for users across the network. Massive MIMO (mMIMO)~\cite{marzetta2016} considers the scenario where there are a large number of antennas serving few users. This concept is coupled with cell-free networks, and branded as cell-free mMIMO, is commonly studied together in the literature~\cite{ngo2017,bjornson2020mmse}. There is rapidly growing interest in this technology and it is seen as one of the drivers for future mobile networks~\cite{ammar2022ucsurvey,zhang2020b5g}. This motivates us to study the resource allocation problem for user-centric cell-free networks.

The authors in~\cite{atzeni2021,kaviani2012} studied linear beamforming strategies to maximize the sum rate and minimize the weighted mean square error (MSE) whereas~\cite{masoumi2020,Interdonato2020} studied power control strategies using geometric programming. Power allocation is a sub-problem of beamforming as it can be implicitly optimized using beamforming weights. The above works focus on beamforming in cell-free mMIMO networks but do not take user scheduling into account.

In contrast, authors of~\cite{dandrea2021} studied the user-AP association problem using the Hungarian algorithm under the cell-centric cell-free setup and~\cite{nguyen2020,dong2019} studied the joint user association and power allocation problem optimizing for energy efficiency. The above works considered the user scheduling problem only for cell-free networks but retained the cellular structure. In a cell-centric cell-free network, the users maybe served by different disjoint clusters of APs but the cellular structure still remains for the APs. Compared to the user-centric clustering, this approach does not provide substantial gains in performance since the inter-cluster interference is not scalable with respect to the intra-cluster signals~\cite{lozano2013}. More importantly, the cell-edge users still receive weak signals~\cite{zhu2016}.

Despite the performance advantage user-centric clustering brings, it makes resource allocation, specifically user scheduling, very difficult since each user may have its own serving cluster of APs. The authors of~\cite{Demir2021} studied receive beamforming for user-centric mMIMO with significantly more antennas than users. The results show that serving all users at max power maximizes sum SE because interference cancellation is possible using the great number of antennas. They also studied the centralized and distributed operation modes, drawing attention to the problem of scalability of cell-free networks. This work built a good foundation for modelling user-centric cell-free networks. However, it is more realistic to consider a dense population of users where spatial resources are scarce. Even with a large number of antennas, the increase in number of users precludes the use of ``massive'' MIMO, reverting to the use of ``MIMO''. Previous work has optimized resource allocation to maximize the WSR in the \textit{downlink} of cell-free MIMO networks~\cite{ammar2022,Ammar2022distributed}.

In the case where there are not enough resources for all users in the network, the authors of~\cite{ammar2022} jointly optimized user scheduling and beamforming by maximizing WSR with similar techniques as used in~\cite{shen2018}. However, in the user-centric cell-free case, the fact that the utility functions are coupled precludes the use of the Hungarian algorithm for scheduling; we require combinatorial search methods. As an alternative,~\cite{ammar2022} used compressive sensing to optimize user scheduling and formulated an iterative algorithm that converges smoothly to a local optimum of the WSR maximization problem.

Scalability issues arise when both the number of users and APs increase. It is not practical to burden a single CPU for all computation tasks in the network. Thus, the distributed operation mode, where APs can process tasks locally, is more practical~\cite{bjornson2020mmse}. Such an approach may provide worse performance because it lacks coordination amongst APs and a global view of the network~\cite{interdonato2019}. This motivates us to add an additional layer of processing units between the CPU and APs such that joint processing can be done among the APs within the same region. By introducing coordination, performance is improved while maintaining reasonable scalability. The downlink resource allocation in this new network was first studied in~\cite{Ammar2022distributed} as an extension of~\cite{ammar2022}. It proposed a decentralized resource allocation algorithm based on a hybrid leakage and interference metric proposed in~\cite{li2022,gamvrelis2022}.

\subsection{Contribution}
As is clear, the only available works in resource allocation for user-centric, cell-free, MIMO settings has addressed the (simpler) downlink, leaving the uplink case untouched. This paper helps fill this gap; specifically the contribution of this paper is as follows:
\begin{enumerate}
	\item We develop resource allocation algorithms and formulating corresponding mathematical models in an uplink user-centric cell-free setting for three different network operation schemes referred as centralized, distributed, and semi-distributed in this paper. The APs take different roles in the signal transmission process for each scheme resulting different advantages and drawbacks. The resource allocation algorithm performs user scheduling and power allocation, maximizing a SINR-based metric, using FP and compressive sensing.
	\item We propose a metric similar to SINR but only includes partial interference and develop decentralized resource allocation algorithms using this proposed metric. These approaches are fully decentralized and do not require any real-time information exchange between APs while providing comparable performance.
	\item We test and compare the combinations of the above three network operation modes and two (proposed metric and SINR-based metric) algorithmic approaches. Our results highlight the trade-offs between performance and overhead costs using different schemes.
\end{enumerate}

\subsection{Organization and Notation}
The rest of the paper is organized as follows. Section~\ref{chapter:model} presents the network and system model for centralized and distributed operation modes. Section~\ref{chapter:centralize_mode} formulates the resource allocation problem for the centralized operation scheme and develops steps to solve this problem. Section~\ref{chapter:distributed_mode} casts the problem in distributed and semi-distributed operation modes and develops an approach for resource allocation. Section~\ref{chapter:decentralize_mode} proposes a new metric for the resource allocation to achieve decentralization. Section {\ref{section:system_comparisons}} compares five proposed systems regarding operation and scalability. Section~\ref{chapter:results} reports on numerical results and findings. Finally, Section~\ref{chapter:conclusion} concludes the discussion.

Our notation is as follows: scalars are represented with lower-case letters (e.g., $a$) and vectors and matrices are represented with lower and upper case bold letters (e.g., $\mathbf{a}$ and $\mathbf{A}$) respectively. The operators $(\cdot)^{-1}$, $(\cdot)^{T}$, and $(\cdot)^{H}$ denote inverse, transpose, and conjugate transpose of a matrix. The $\ell_p$ norm is denoted as $||\cdot||_p$ and $\mathbf{I}_m$ is the $m$-dimensional identity matrix. Calligraphic letters (e.g., $\mathcal{A}$) indicate sets. Complex matrices of size $M \times N$ is denoted with $\mathbb{C}^{M \times N}$ and $N=1$ for column vectors. A complex Gaussian random vector $\mathbf{x}$ is denoted as $\mathbf{x} \sim \mathcal{CN}(\mathbf{m},\mathbf{R})$ with mean $\mathbf{m}$ and covariance matrix $\mathbf{R}$.

\section{System Model}\label{chapter:model}

\subsection{Network Model}\label{subsection:network_model}
\begin{figure}[t]
	\centering
	\includegraphics[scale=0.5]{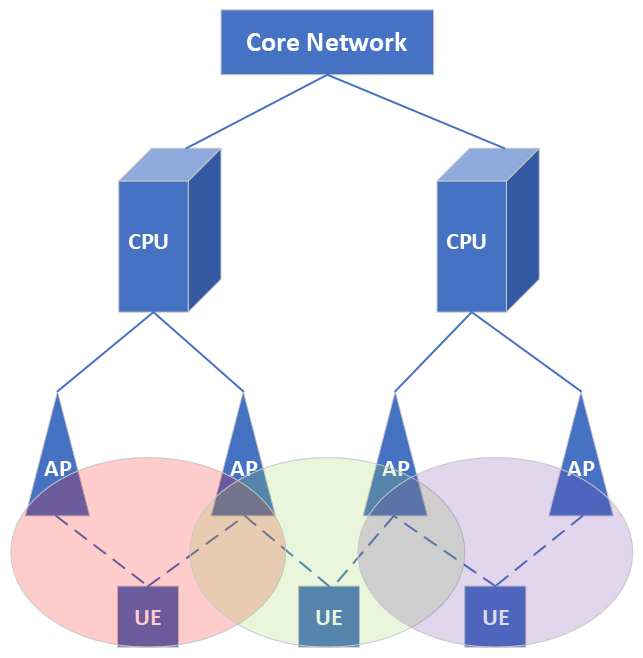}
	\caption{ Architecture of a user-centric cell-free network.}
	\label{fig:network_model}
\end{figure}

We consider a user-centric cell-free MIMO network, which operates in time-division duplex (TDD) mode. Fig.~\ref{fig:network_model} shows an example of user-centric cell-free network model where the dashed lines represent wireless links and solid lines represent wired links. Our model comprises $Q$ central processing units (CPUs) each controlling a disjoint set of $R$ APs for a total of $QR$ APs; the set of APs connected to CPU $q$ is denoted by $\mathcal{B}_q$, while $\mathcal{B}$ denotes the set of all APs. Each AP is equipped with $M$ antennas while each user equipped with $N$ antennas. The set $\mathcal{U}$, with size $\mid \mathcal{U} \mid$, denotes the set of all users or user equipments (UEs). In our model, we expect $\mid \mathcal{U} \mid \gg QR$, i.e., the number of users exceeds the available resources.

We use indices $q$, $r$, $u$ to refer to CPUs, APs, and users respectively throughout the paper. We consider a block fading channel model, i.e., a relatively low-mobility scenario. We model the uplink channel between user $u$ and AP $r$ $\mathbf{H}_{ru} \in \mathbb{C}^{M \times N}$ as $\sqrt{\psi_{ru} \beta(d_{ru})} \mathbf{G}_{ru}$, where $\mathbf{G}_{ru}$ accounts for small-scale fading and is modeled as Rayleigh with each entry having unit variance. The terms $\psi_{ru}$ and $\beta(d_{ru})$ are denote the large-scale fading (modeled as log-normal shadowing) and pathloss respectively; here, $d_{ru}$ denotes the distance between AP $r$ and user $u$. The path loss is modeled as $\beta(d_{ru}) \propto d_{ru}^{-\alpha}$, where $\alpha$ is the pathloss exponent.

Based on the channel model, for our user-centric setup, for each user $u \in \mathcal{U}$, we define its serving cluster $\mathcal{C}_u$ comprised of nearby APs. It is important to  note that the serving cluster represents the set of all APs that can \textit{potentially} serve the user; whether such a link is in fact used depends on scheduling decisions, as we will explain later. Our criterion for the serving cluster is $\mathcal{C}_u = \{ r \in \mathcal{B} :\psi_{ru} \beta(d_{ru}) \ge \rho \} \cup \{ r \in \mathcal{B}: \mathrm{arg \: max}_r \: \psi_{ru} \beta(d_{ru}) \} $. The first set of the union eliminates all APs that do not contribute a useful signal to the user, while the second set ensures all users have at least one AP in its serving cluster. 

We define the set $\mathcal{E}_r$ as the set of users that can be served by AP $r$; this set can be obtained from $\mathcal{C}_u, u = 1, \dots, \mid\mathcal{U}\mid$. We note that, for a particular user, the APs in its serving cluster may be controlled by different CPUs. Specifically, given user-centric clustering, different sets $\mathcal{E}_r$ and $\mathcal{E}_{r'}$ can overlap for $r \neq r'$ because a user can belong to multiple APs, or, similarly, multiple CPUs.

In the latter sections, we will introduce different uplink network operation modes that estimate the channel at different locations (APs or CPU). However, since the channel vectors are independent, there is no loss in optimality if the channel estimates are computed separately at each AP or together at the CPU. In~\cite{Ammar2022distributed}, we developed channel estimation techniques for our network model. There, we propose a heuristic of a low-overhead pilot assignment policy that assigns the same pilots to users that are as far from each other as possible. The users then transmit their pilots and the channels are estimated at APs or CPUs (depending on operation mode) with linear minimum MSE (MMSE) estimator. Thus, channel estimation will not be the focus for this paper and we assume perfect CSI.

\subsection{Model for Centralized Operation}\label{subsection:centralized_model}
In this section, we consider the case of centralized operation where there is only one CPU in the network. All APs serve as remote radio heads (RRH) that forward the received signals to the CPU for processing. 

For the signal model at each symbol period, in the uplink, the signal received at each AP $r$ is given by
\begin{equation} \label{signal_model_centralized}
	\mathbf{y}_r = \sum_{u' \in \mathcal{U}} \mathbf{H}_{ru'} \mathbf{S}_{u'} \mathbf{v}_{u'} x_{u'} 
	+ \mathbf{z}_r
\end{equation}
where $\mathbf{y}_r \in \mathbb{C}^{M \times 1}$ is the received signal received at AP $r$, $\mathbf{S}_u=s_u \mathbf{I}_N$ is the diagonal matrix created using the binary scheduling variable $s_u$ indicating whether user $u$ is scheduled or not. Due to the broadcast assumption, if user is scheduled, then it is scheduled on all APs of its serving cluster. Similarly, $\mathbf{v}_u \in \mathbb{C}^{N \times 1}$ is the transmit beamformer for user $u$ and it is common to all APs $r \in \mathcal{C}_u$ due to the broadcast nature of the uplink. Finally, $x_u$ is symbol sent by user $u$ and $\mathbf{z}_r \in \mathbb{C}^{M \times 1}$ is the noise received. Note that we consider single-streams to users despite the fact that they are equipped with multiple antennas. The extra antennas are used to suppress interference rather than to spatially multiplex multiple streams. We will further discuss and justify this choice in Section~\ref{chapter:results}.

The received uplink signal at all APs are jointly used at the CPU to estimate the symbol transmitted by user $u$,
\begin{align} 
	\hat{x}_u &= \sum_{r \in \mathcal{C}_u} \mathbf{w}_{ru}^H \mathbf{y}_r \nonumber \\
	&= \mathbf{w}_{u}^H ( \sum_{u' \in \mathcal{U}} \mathbf{H}_{u,u'} \mathbf{S}_{u'}           \mathbf{v}_{u'} x_{u'} + \mathbf{z}_u ) \label{estimate_symbol_centralized}
\end{align}
where $\mathbf{w}_{ru} \in \mathbb{C}^{M \times 1}$ is the receive beamformer at AP $r \in \mathcal{C}_u$ for user $u$. \eqref{estimate_symbol_centralized} substitutes the received signal with {\eqref{signal_model_centralized}}. The terms are rearranged and combined with $x_{u'}$ as the common terms and uses a new set of vectorized variables to simplify the double summation into a single sum. The vectors $\mathbf{w}_{u} \in \mathbb{C}^{M|\mathcal{C}_u| \times 1}$, $\mathbf{z}_u \in \mathbb{C}^{M|\mathcal{C}_u| \times 1}$, $\mathbf{H}_{u,u'} \in \mathbb{C}^{M|\mathcal{C}_u| \times N}$ are the vertically concatenated receive beamformers, noise, and channels respectively. For instance, $\mathbf{H}_{u,u'}$ is the concatenation of channels $\{ \mathbf{H}_{ru'}: r \in \mathcal{C}_u \}$ between cluster $\mathcal{C}_u$ serving $u$ and user $u'$ or equivalently $\mathbf{H}_{u,u'} = [ 
\mathbf{H}_{r_1u'}^T \mathbf{H}_{r_2u'}^T \ldots \mathbf{H}_{r_{|\mathcal{C}_u|}u'}^T ]^T$. Similarly, $\mathbf{w}_{u}$ is the concatenation of receive beamformers $\{ w_{ru}: r \in \mathcal{C}_u \}$ and is used at the CPU to estimate the symbol sent by user $u$ or equivalently $\mathbf{w}_u = [ \mathbf{w}_{r_1u}^T \mathbf{w}_{r_2u}^T \ldots \mathbf{w}_{r_{|\mathcal{C}_u|}u}^T ]^T $.

Based on~\eqref{estimate_symbol_centralized}, we can derive the SINR for user $u$ as
\begin{equation}\label{sinr_centralized}
	\mathrm{SINR}_u = \frac{| \mathbf{w}_{u}^H \mathbf{H}_{u,u} \mathbf{S}_{u} \mathbf{v}_{u} |^2}
	{ \sigma^2 || \mathbf{w}_{u} ||^2 + 
		\sum_{u' \in \mathcal{U} \backslash u} | \mathbf{w}_{u}^H \mathbf{H}_{u,u'} \mathbf{S}_{u'} \mathbf{v}_{u'} |^2  }.
\end{equation}
Furthermore, using the MMSE receiver for $\mathbf{w}_{u}$, given by,
\begin{equation}
	\mathbf{w}_{u}^\mathrm{MMSE} = 
    ( \sigma^2\mathbf{I}_{|\mathcal{C}_u|M} 
	+\sum_{u' \in \mathcal{U}} \mathbf{H}_{u,u'} \mathbf{v}_{u'} \mathbf{v}_{u'}^H \mathbf{H}_{u,u'}^H ) ^{-1}
	\mathbf{H}_{u,u} \mathbf{v}_{u} \label{eq:MMSEreceiver_centralized}
\end{equation}
the SINR expression can be further reduced to 
\begin{align}
	\mathrm{SINR}_u & = \mathbf{v}_{u}^H \mathbf{S}_{u}^H \mathbf{H}_{u,u}^H \times \nonumber \\
	& \hspace*{-0.3in}
	\left( \sigma^2 \mathbf{I}_{M} + \sum_{u' \in \mathcal{U} \backslash u} \mathbf{H}_{u,u'} \mathbf{S}_{u'} \mathbf{v}_{u'} \mathbf{v}_{u'}^H \mathbf{S}_{u'}^H \mathbf{H}_{u,u'}^H \right)^{-1} 
	\mathbf{H}_{u,u}  \mathbf{S}_{u} \mathbf{v}_{u} \label{sinr_centralized_multidim}
\end{align}
Note that this SINR reformulation is just for simplicity does not change the outcome of resource allocation algorithm. Later we will see that the receive beamformer terms reappear in another form during the algorithm.

Having developed the signal model for the centralized case, it is worth noting why alternative operation modes are required. In the centralized case, as seen in~\eqref{estimate_symbol_centralized} and~\eqref{eq:MMSEreceiver_centralized}, we see that the CPU requires both global CSI and the fronthaul must return, to the APs, the beamforming weights, all within a reasonable fraction of the channel coherence time. This presents a significant signaling overhead and associated problems like fronthaul quantization distortion~\cite{Demir2021}. This leads us to consider the second, distributed, operation mode.

\subsection{Model for Distributed Operation}\label{subsection:distributed_model}
In the distributed operation mode, APs no longer serve as a relay to the CPUs. Each AP is equipped with a baseband processor and is capable of channel estimation and makes scheduling and receive beamforming decisions locally. This significantly reduces the overload for channel estimation and signaling in the fronthaul.

In the first stage, the signal received at each AP is the same as in the centralized case, as given in~\eqref{signal_model_centralized}. Instead of passing the signal to an upper processing unit, at AP $r$, the estimation of symbol for user $u$ is done locally using 
\begin{align} \label{estimate_symbol_distributed}
	& \hat{x}_{ru} = \mathbf{w}_{ru}^H \mathbf{y}_r \nonumber \\
	&= \mathbf{w}_{ru}^H \mathbf{H}_{ru} \mathbf{S}_{u}\mathbf{v}_{u} x_{u} + \sum_{u' \in \mathcal{U} \backslash u} \mathbf{w}_{ru}^H \mathbf{H}_{ru'} \mathbf{S}_{u'} \mathbf{v}_{u'} x_{u'} + \mathbf{w}_{ru}^H \mathbf{n}_r 
\end{align}
where $\mathbf{w}_{ru}$ is the local receive beamformer and $\mathbf{n}_r$ is the noise.

In the second stage, the local data estimates of all APs are gathered at the CPU and combined into a final estimate of user data. At the CPU, the weighted combined estimate of each user symbol is
\begin{equation}
	\hat{x}_{u} = \sum_{r\in \mathcal{C}_u} a_{ru} \hat{x}_{ru} \label{eq:signal_combined_distributed}
\end{equation}
where $a_{ru}$ is the combination weight for symbol from user $u$ estimated at and passed by AP $r$. Using~\eqref{estimate_symbol_distributed}, the expression in~\eqref{eq:signal_combined_distributed} comprises three terms: the first is the signal component, the second captures the inter-user interference, and the third the noise. Based on this expression, with more compact notation, we can derive the SINR for user $u$ as
\begin{equation}\label{sinr_distributed}
	\mathrm{SINR}_u = \frac{| \mathbf{a}_{u}^H \mathbf{g}_{u,u} |^2}
	{ \mathbf{a}_{u}^H ( \mathbf{F}_{u} + 
		\sum_{u' \in \mathcal{U} \backslash u} | \mathbf{a}_{u}^H \mathbf{g}_{u,u'} |^2 ) \mathbf{a}_{u} } 
\end{equation}
where $\mathbf{a}_u \in \mathbb{C}^{|\mathcal{C}_u| \times 1}$ is the concatenation of weights $\{ \mathbf{a}_{ru}: r \in \mathcal{C}_u \}$ with $\mathbf{a}_{ru}=0$ for $r \notin \mathcal{C}_u$, $\mathbf{F}_u = diag\{ \sigma^2 || \mathbf{w}_{ru} ||^2: r \in \mathcal{C}_u \} \in \mathbb{C}^{|\mathcal{C}_u| \times |\mathcal{C}_u|}$ is the diagonal noise matrix, and $\mathbf{g}_{u,u'} \in \mathbb{C}^{|\mathcal{C}_u| \times 1}$ is the concatenated interference vector $\{ \mathbf{w}_{ru}^H \mathbf{H}_{ru'} \mathbf{S}_{u'} \mathbf{v}_{u'}: r \in \mathcal{C}_u \}$ representing the interference contribution from user $u'$ to user $u$ during local estimation at AP $r \in \mathcal{C}_u$ (setting $u' = u$ provides the signal term in the numerator).

In the following two sections, we develop resource allocation algorithms for centralized, distributed and semi-distributed operations. It is worth emphasizing that such algorithms have not been developed for the \textit{uplink}. As we will see, the algorithms in Sections~\ref{chapter:centralize_mode} and~\ref{chapter:distributed_mode} require substantial knowledge of the state of the network and are, thereby, impractical. However, the more practical algorithm in Section~\ref{chapter:decentralize_mode} builds on these initial attempts. 

\section{Centralized Operation}\label{chapter:centralize_mode}

\subsection{Problem Definition}
Given the system models in the previous section, we can now formulate the WSR optimization problem over the transmit beamformers and user scheduling variables (we use the MMSE receive beamformers as in~\eqref{eq:MMSEreceiver_centralized}). Let $\delta_u$ denote the weight for user $u$ and $s_u$ denote whether user $u$ is scheduled ($s_u = 1$) or not ($=0$). Let $\mathcal{S}$ denote the set of scheduling variables and $\mathcal{V}$ denote the set of transmit beamformers, as defined earlier. 
\begin{subequations}\label{eqn:optimization_problem_centralized}
	\begin{align}
		\max_{ \mathcal{S}, \mathcal{V}}\quad & \displaystyle\sum\limits_{u \in \mathcal{U}}
		\delta_{u} \log\left( 1 + 
		\mathrm{SINR}_{u}
		\right) ,
		\label{subeqn:obj_func}
		\\
		\text{s.t.}\quad 
		& s_{u} \in \{0,1\} \,\, \forall u \in \mathcal{U},
		\label{subeqn:cnstr_sched_dom}
		\\
		&\displaystyle\sum\limits_{u \in \mathcal{U}} s_{u} \le RM,
		\label{subeqn:cnstr_max_sched}
		\\
		& ||\mathbf{v}_{u}||^2 \le P_T \,\,\forall u \in \mathcal{U}.
		\label{subeqn:cnstr_power}
	\end{align}
\end{subequations}
The objective function~\eqref{subeqn:obj_func} denotes the weighted sum of user spectral efficiencies where the term $\mathrm{SINR}_u$ is defined in~\eqref{sinr_centralized_multidim} and $\delta_{u}$ as the weight for user $u$. The first set of constraints in~\eqref{subeqn:cnstr_sched_dom} enforce binary scheduling decisions. The second set of constraints in~\eqref{subeqn:cnstr_max_sched} is the capacity constraint to ensure the number of users scheduled does not exceed the number of antennas in the system. In the centralized operation case, there is only one CPU so $Q=1$ and there is a total of $R$ APs in the network; thus, a total of $RM$ antennas are available. Constraints~\eqref{subeqn:cnstr_power} impose a limit of $P_T$ on the power available  at each user for transmission. 

It is worth noting that similar WSR resource allocation problems have been proven to be NP-hard~\cite{luo2008}. Hence, obtaining a global optimum is mostly computationally intractable. 

\subsection{Problem Analysis}
In this section, we analyze the problem and reduce it to a decoupled and linear form which will help us developing the resource allocation algorithm in the next section. 

In~\cite{li2022} and~\cite{shen2018}, the authors studied similar WSR maximization problems but in cellular systems. They handled the user scheduling optimization with a combinatorial search algorithm where non-zero beams are tried on different users to find an optimal scheduling that maximize the network utility on a per-cell basis. Unfortunately, we cannot follow the same procedure in the cell-free case because the problem is coupled between APs due to user-centric clustering.

The authors of~\cite{ammar2022} had encountered the similar problem in the downlink. To find an alternative for the combinatorial search, they had reformulated the problem so that the discrete scheduling variable is removed from the equations. In summary, the solution requires transforming the discrete constraints in~\eqref{subeqn:cnstr_sched_dom} and~\eqref{subeqn:cnstr_max_sched} into a continuous weighted norm constraint with the technique of weighted $l1$-norm minimization from compressive sensing. The reformulated optimization problem becomes
\begin{subequations}\label{eqn:optimization_problem_centralized_reform}
	\begin{align}
		\max_{ \mathcal{V}}\quad & \displaystyle\sum\limits_{u \in \mathcal{U}}
		\delta_{u} \log\left( 1 + 
		\gamma_{u}
		\right) 
		\label{subeqn:obj_func_reform}
		\\
		\text{s.t.}\quad 
		& \gamma_{u} = \mathbf{v}_{u}^H \mathbf{H}_{u,u}^H
		\Bigl( \sigma^2 \mathbf{I}_{|\mathcal{C}_u|M} +  \nonumber \\
		& \hspace*{0.5in} \sum_{u' \in \mathcal{U} \backslash u} \mathbf{H}_{u,u'}  \mathbf{v}_{u'} \mathbf{v}_{u'}^H \mathbf{H}_{u,u'}^H \Bigl)^{-1} 
		\mathbf{H}_{u,u}  \mathbf{v}_{u}
		\label{subeqn:sinr_gamma_reform}
		\\
		&\displaystyle\sum\limits_{u \in \mathcal{U}} \alpha_u ||\mathbf{v}_{u}||^2 \le RM
		\label{subeqn:cnstr_max_sched_reform}
		\\
		& ||\mathbf{v}_{u}||^2 \le P_T
		\label{subeqn:cnstr_power_reform}
	\end{align}
\end{subequations}
where~\eqref{subeqn:sinr_gamma_reform} is similar to SINR but with the scheduling variable removed and~\eqref{subeqn:cnstr_max_sched_reform} is the reformulated constraint.

\subsection{Resource Allocation for Centralized Operation}\label{subsection:recourse_allocation_cent}
We employ an iterative optimization approach based on fractional programming (FP)~\cite{shen2018} to find a fixed point for the WSR maximization problem in~\eqref{eqn:optimization_problem_centralized_reform}. FP offers a set of tools such as Lagrange Dual Transform and Quadratic Transform that can reformulate the original objective function~\eqref{subeqn:obj_func_reform} that is in a tedious form of sum of logarithmic ratio into a linear problem. 

To reduce the coupling between the ratio SINR term and the logarithmic function, we first apply Lagrangian Dual Transform to the objective function to obtain 
\begin{equation}
	\begin{split}
		f_r(\mathcal{V},\Gamma) = \sum_{u \in \mathcal{U}} \delta_{u} \Bigl( \log (1+\gamma_{u}) - \gamma_{u} +  (1+\gamma_{u})
		\mathbf{v}_{u}^H \mathbf{H}_{u,u}^H 
        \\ \times ( \sigma^2\mathbf{I}_{|\mathcal{C}_u|M} 
		+\sum_{u' \in \mathcal{U}} \mathbf{H}_{u,u'} \mathbf{v}_{u'} \mathbf{v}_{u'}^H \mathbf{H}_{u,u'}^H 
		) ^{-1}
		\mathbf{H}_{u,u} \mathbf{v}_{u} \Bigl)
	\end{split}
\end{equation}
where $\Gamma$ represents the set of auxiliary SINR variables $\gamma_u, u\in\mathcal{U}$. To find the optimal value for each variable in the objective function, we use block coordinate descent where we solve for optimal value for each set of variables while holding other variables fixed. Having $\mathcal{V}$ held fixed, we can the optimal $\gamma_{u}$ by solving $\partial f_r/\partial\gamma_{u}=0$ with
\begin{equation}\label{eqn:optimal_gamma_cent}
\begin{split}
	\gamma_{u}^*=\mathbf{v}_{u}^H \mathbf{H}_{u,u}^H ( \sigma^2\mathbf{I}_{|\mathcal{C}_u|M} + 
	\sum_{u' \in \mathcal{U} \backslash u} \mathbf{H}_{u,u'} \mathbf{v}_{u'} \mathbf{v}_{u'}^H \mathbf{H}_{u,u'}^H ) ^{-1}
	\mathbf{H}_{u,u} \mathbf{v}_{u}
 \end{split}
\end{equation}

The reformulated objective function $f_r$ is in a sum of ratios form which still remains non-linear and coupled, we use Quadratic Transform to further reformulate the objective function obtaining
\begin{equation}
\begin{split}
    f_q(\mathcal{V},\Gamma,\mathcal{Y}) = 
    \sum_{u \in \mathcal{U}} \delta_u (\log (1+\gamma_{u}) - \gamma_{u}) \\
    & \hspace*{-2.in}+ \sum_{u \in \mathcal{U}} \Bigl( 2 \sqrt{\delta_{u}(1+\gamma_{u})} 
    \mathrm{Re}\{ \mathbf{v}_{u}^H \mathbf{H}_{u,u}^H \mathbf{y}_{u} \}  \\
    & \hspace*{-2.in}- \mathbf{y}_{u}^H 
    ( \sigma^2\mathbf{I}_{|\mathcal{C}_u|M} + \sum_{u' \in \mathcal{U}} \mathbf{H}_{u,u'} \mathbf{v}_{u'} \mathbf{v}_{u'}^H \mathbf{H}_{u,u'}^H )
    \mathbf{y}_{u} \Bigl)
\end{split}
\end{equation}
where $\mathcal{Y}$ represents the set of auxiliary variables $\mathbf{y}_u \in \mathbb{C}^{M \times 1}$ introduced by the Quadratic transform. As before, we can obtain the optimal auxiliary variable in $\mathcal{Y}$ by holding the other variables constant and solving for $\partial f_q/\partial\mathbf{y}_{u}=0$; as a result, we obtain
\begin{equation}\label{eqn:optimal_y_cent}
	\begin{split}
	\mathbf{y}_{u}^* & =  \sqrt{\delta_{u}(1+\gamma_{u})}
	\Bigl( \sigma^2\mathbf{I}_{|\mathcal{C}_u|M} \\
	& \hspace*{0.3in} +\sum_{u' \in \mathcal{U}} \mathbf{H}_{u,u'} \mathbf{v}_{u'} \mathbf{v}_{u'}^H \mathbf{H}_{u,u'}^H 
	\Bigl) ^{-1}
	\mathbf{H}_{u,u} \mathbf{v}_{u}
	\end{split}
\end{equation}
Observe that the optimal $\mathbf{y}_{u}$ is exactly the MMSE receiver scaled by a factor. 

The transmit beamformers $\mathbf{v}_{u}$ are constrained by~\eqref{subeqn:cnstr_max_sched_reform} and~\eqref{subeqn:cnstr_power_reform}. The Lagragian formulation of $f_q$ becomes
\begin{equation}
	\begin{split}
		f_L(\mathcal{V},\Gamma,\mathcal{Y}) & = f_q + \sum_{u \in \mathcal{U}} \mu_u (P_T - ||\mathbf{v}_{u}||^2) \\
		& + \lambda \sum_{u \in \mathcal{U}} (RM - \alpha_u ||\mathbf{v}_{u}||^2)
	\end{split}
\end{equation}
where $\lambda$ and $\mu_u$ are the Lagrange multipliers. We then can find the optimal transmit beamformer for $\mathcal{V}$ by fixing other variables and solving for $\partial f_L/\partial\mathbf{v}_{u}=0$. The optimal $\mathbf{v}_{u}$ equals to 
\begin{eqnarray}
	& \mathbf{v}_{u}^* =
	\sqrt{\delta_{u}(1+\gamma_{u})}  \Bigl( (\lambda^* \alpha_u + \mu_u^*) \mathbf{I}_N \nonumber \\
	& 
	\hspace*{0.5in} + \sum_{u' \in \mathcal{U}} \mathbf{H}_{u',u}^H \mathbf{y}_{u'} 
	\mathbf{y}_{u'}^H \mathbf{H}_{u',u} \Bigl)^{-1}
	\mathbf{H}_{u,u}^H \mathbf{y}_{u} \label{eqn:optimal_v_cent}
\end{eqnarray}
The Lagrange multipliers are non-negative real numbers and can be determined through the power and reformulated capacity constraints in~\eqref{subeqn:cnstr_max_sched_reform} and~\eqref{subeqn:cnstr_power_reform}. 

It is important to note that both of these constraints depended on the user power. By complementary slackness, one of the Lagrange multiplier must be zero or only one constraint remains tight; however, we do not know in advance which constraint is relaxed. The authors in~\cite{ammar2022} propose the heuristics that first checks if the capacity constraint is satisfied or not with $\lambda=0$; if not satisfied, $\lambda$ is set to a small value and we find $\mu_u$ with bisection search to satisfy the power constraint. From our tests, this procedure works well and $\lambda$ always converges to zero after a few iterations of the algorithm.

Both FP and weighted $\ell 1$-norm minimization~\cite{candes2007} are iterative so we can update the weights $\alpha_u$ in~\eqref{subeqn:cnstr_max_sched_reform} simultaneously. We construct an iterative process to find the weights with
\begin{equation}\label{eqn:optimal_alpha_cent}
	\alpha_u^{(j+1)} = \frac{1}{||\mathbf{v}_{u}^{(j)}||^2 + \epsilon}
\end{equation}
where $\mathbf{v}_{u}^{(j)}$ is the transmit beamformer for user $u$ at the current iteration and $\alpha_u^{(j+1)}$ is the weight used in next iteration. Generally, the value of $\epsilon$ is chosen slightly below the expected value of $||\mathbf{v}_{u}||^2$ for the scheduled users~\cite{candes2007}. We denote $\boldsymbol{\alpha}$ as the set of $\alpha_u$ for all users.

The role of the weights $\boldsymbol{\alpha}$ is crucial and it is worth considering optimizing this weight achieves the purpose of user scheduling. Typically in a resource allocation problem, the desired users are the users with high SINR and, also, causing minimal interference to other active users. These users are mostly a solution to a local optimum and remain at high power while other users' power decrease over iterations. As seen in~\eqref{eqn:optimal_alpha_cent}, the weight $\alpha_u$, for user $u$, is inversely proportional to the user's transmit power. Thus if the capacity constraint is not satisfied for an iteration, $\lambda \neq 0$ and the weight $\alpha_u$ starts to play a role in the optimal transmit beamformer equation~\eqref{eqn:optimal_v_cent} with an inverse relationship. As a result, the undesired users with large weights transmit with even lower power. Over multiple iterations, this power will converge to zero (equivalently, these users are not scheduled), so that the capacity constraint is met. Since weighted $l1$-norm minimization arises from compressive sensing, we can also interpret as the procedures as finding the optimal scheduling variables by exploiting its sparsity.

Given the expressions for the optimal variables, we can now define the iterative algorithm for resource optimization in the centralized case. Algorithm~\ref{alg:resource_allocation_centralized} details the proposed approach. It initializes all users to transmit at full power, initializes the weights, and then iteratively updates $\Gamma$ (the SINR), $\mathcal{Y}$ (proportional to the MMSE receiver), $\mathcal{V}$ (the transmit beamformer), $\boldsymbol{\alpha}$ (the weight in the reformulation of the capacity constraint). 
\begin{algorithm}
	\caption{Resource Allocation Algorithm for Centralized Operation}\label{alg:resource_allocation_centralized}
	\begin{algorithmic}[1]
		\State{Initialize $\mathbf{v}_{u}$ for all users such that $||\mathbf{v}_{u}||^2=P_T $}\label{step:initialization_1}
		\State{Initialize $\alpha_{u}=1/P_T$ for all users}\label{step:initialization_2}
		\State{\textbf{repeat}}
		\State{\quad Update $\Gamma$ using~\eqref{eqn:optimal_gamma_cent}}
		\State{\quad Update $\mathcal{Y}$ using~\eqref{eqn:optimal_y_cent}}
		\State{\quad Update $\mathcal{V}$ using~\eqref{eqn:optimal_v_cent}}
		\State{\quad Update $\boldsymbol{\alpha}$ using~\eqref{eqn:optimal_alpha_cent}}
		\State {${\mathbf{until}}$ convergence }
	\end{algorithmic}
\end{algorithm}

In practice, the CPU only has imperfect CSI. In this case, the estimated channels replace the true channels in the algorithm. For robustness, the covariance matrices of estimation error are included in the denominator of the SINR expression. We had developed FP-based algorithms under imperfect CSI in~\cite{ammar2022,Ammar2022distributed,li2022,gamvrelis2022} with only small degradation in system performance.


\section{Distributed and Semi-Distributed Operation}\label{chapter:distributed_mode}

In the previous section, we studied the resource allocation problem under centralized operation. Under the centralized network architecture, the APs serve as RRHs and forward the received signal to the CPU that does all the channel estimation, receive combining, and data detection. As mentioned, this results in significant overhead in acquiring CSI and fronthaul signaling which must carry both pilot signals and data. There is also a computation scalability issue with calculating the optimal MMSE receiver vector~\cite{Björnson2017}. In addition, resource allocation with FP-based algorithm has computational complexity scales with total number of users in the network which further worsens the issue of scalability. In this section, we find alternative solutions to address these practical issues by changing the network operation architecture.

\subsection{Resource Allocation for Distributed Operation}\label{subsection:resource_alloc_distr}
The model for distributed network is defined in Section~\ref{subsection:distributed_model} where each AP is equipped with a baseband processing unit and is capable of channel estimation, receive combining, and making scheduling decisions locally. In other words, the CPU distributes out the signal processing tasks to its APs.

We define the WSR optimization problem over transmit beamformers and user scheduling variables in a similar fashion as~\eqref{eqn:optimization_problem_centralized}. However, we do not replace the SINR term with the distributed SINR expression in~\eqref{sinr_distributed} because it would represent a completely different operation mode where channel estimation and receive combining is done at the AP but scheduling decisions are still done at the CPU. Since scheduling would still be centralized, we do not consider this situation in this paper, but is an interesting avenue for future research. Nonetheless, the distributed SINR expression in~\eqref{sinr_distributed} serves the purpose of measuring the final system performance but does not take part in resource allocation.

To develop the operation scheme where CPU distributes the resource allocation tasks to the APs, we first define a \textit{psuedo-metric} $\mathrm{SINR}_{ru}$ that represents the \textit{local} SINR for user $u$ with respect to AP $r$
\begin{equation}\label{eqn:sinr_distr}
	\mathrm{SINR}_{ru} = \boldsymbol{\tau}_{ru}^H \mathbf{H}_{ru}^H
	( \sigma^2 \mathbf{I}_{M} + \sum_{u' \in \mathcal{U} \backslash u} \mathbf{H}_{ru'}  \mathbf{v}_{u'} \mathbf{v}_{u'}^H \mathbf{H}_{ru'}^H )^{-1} 
	\mathbf{H}_{ru}  \boldsymbol{\tau}_{ru}
\end{equation}
Note that, here, we do not use $\mathbf{H}_{u,u}$ for the channel because we are working with one AP only. Furthermore, the scheduling variable is subsumed into the formulation using the compressive sensing approach as in~\eqref{eqn:optimal_alpha_cent}. Here, $\boldsymbol{\tau}_{ru}$ is the local optimal decision where $\boldsymbol{\tau}_{ru} = \mathbf{v}_{u}$ if user $u$ is scheduled on AP $r$, else $\boldsymbol{\tau}_{ru} = \mathbf{0}$. This variable captures information of both scheduling and transmit beamformer only locally at AP $r$. It is important to emphasize that this is different from the centralized operation mode because, \textit{in the distributed case, not all APs from a user's cluster have to serve that user simultaneously}. For instance, a user that transmits at non-zero power can be considered as interference by APs that are in this user's cluster if it is not scheduled on these APs which can never happen under the centralized mode.

The optimization problem becomes
\begin{subequations}\label{eqn:resource_alloc_problem_distr}
	\begin{align}
		\max_{ \mathcal{V}}\quad & \displaystyle\sum\limits_{r \in \mathcal{B}}\sum\limits_{u \in \mathcal{E}_r}
		\delta_{ru} \log\left( 1 + 
		\gamma_{ru}
		\right) 
		\label{subeqn:obj_func_reform_distr}
		\\
		\text{s.t.}\quad 
		& \gamma_{ru} = \mathrm{SINR}_{ru}
		\\
		&\displaystyle\sum\limits_{u \in \mathcal{E}_r} \alpha_{ru} ||\boldsymbol{\tau}_{ru}||^2 \le M
		\\
		& ||\boldsymbol{\tau}_{ru}||^2 \le P_T
	\end{align}
\end{subequations}
where all variables have an additional index $r$ to represent the optimization problem is solved locally at AP $r$. Even though each AP makes their own power allocation decision at each iteration of the algorithm, information exchange between APs are assumed so it is defined as $\mathbf{v}_{u} = \boldsymbol{\tau}_{r^*u}$ where $r^* = \arg\max_r\{ ||\boldsymbol{\tau}_{ru}|| : r \in \mathcal{B} \}$. Essentially, the user transmits with the maximum of the powers it is allocated by the local AP decisions.

Another way we can view this resource allocation problem is optimizing for sum of $|\mathcal{B}|$ sub-problems. The reason we are maximizing for the sum but not each individual sub-problem is because they are coupled. For instance, if a user is allocated by an AP, it is treated as interference by another AP therefore lower the SINR for other users resulting lower data rate. We need to consider the whole picture thus incorporating the summation over all APs. This is made possible with information exchange between APs. We again stress that optimizing for~\eqref{subeqn:obj_func_reform_distr} does not necessarily maximize the sum SE performance evaluated based on~\eqref{sinr_distributed} but it makes resource allocation decisions possible locally at each AP.

We again employ a FP-based iterative optimization approach to find a optimum for this WSR maximization problem. Since simlar procedures have already been analyzed in Section~\ref{subsection:recourse_allocation_cent}, we will cover the steps with fewer details.

After a series of Lagrange Dual Transform, Quadratic Transform, and partial derivatives, we obtain the following formulas for optimal auxiliary variables and transmit beamformer,
\begin{equation}
	\hspace*{-0.025in} \gamma_{ru}^* = \boldsymbol{\tau}_{ru}^H \mathbf{H}_{ru}^H ( \sigma^2\mathbf{I}_M + 
	\sum_{u' \in \mathcal{U} \backslash u} \mathbf{H}_{ru'} \mathbf{v}_{u'} \mathbf{v}_{u'}^H \mathbf{H}_{ru'}^H ) ^{-1}
	\mathbf{H}_{ru} \boldsymbol{\tau}_{ru} \label{eqn:optimal_gamma_distr}
\end{equation}
\begin{equation}
\begin{split}
	& \hspace*{-0.5in}\mathbf{y}_{ru}^* =  \sqrt{\delta_{ru}(1+\gamma_{ru})} \times 
	\Bigl( \sigma^2\mathbf{I}_M + \mathbf{H}_{ru} \boldsymbol{\tau}_{ru} \boldsymbol{\tau}_{ru}^H \mathbf{H}_{ru}^H
	 \\
	& +\sum_{u' \in \mathcal{U} \backslash u} \mathbf{H}_{ru'} \mathbf{v}_{u'} \mathbf{v}_{u'}^H \mathbf{H}_{ru'}^H 
	\Bigl) ^{-1}
	\mathbf{H}_{ru} \boldsymbol{\tau}_{ru} \label{eqn:optimal_y_distr}
 \end{split}
\end{equation}
\vspace*{-0.05in}
\begin{equation}
\begin{split}
	& \hspace*{-0.32in}\boldsymbol{\tau}_{ru}^* = 
	\sqrt{\delta_{ru}(1+\gamma_{ru})} \times \Bigl( (\lambda_{r}^* \alpha_{ru} + \mu_{ru}^*) \mathbf{I}_N  \\
	& + \sum_{r' \in \mathcal{B}} \sum_{u' \in \mathcal{E}_{r'}} \mathbf{H}_{r'u}^H \mathbf{y}_{r'u'} 
	\mathbf{y}_{r'u'}^H \mathbf{H}_{r'u} \Bigl)^{-1}
	\mathbf{H}_{ru}^H \mathbf{y}_{ru} \label{eqn:optimal_tau_distr}
\end{split}
\end{equation}
\vspace*{-0.15in}
\begin{eqnarray}
& \hspace*{-0.6in} \mathbf{v}^*_{u} = \boldsymbol{\tau}_{r^*u} \mathrm{\ where\ } r^* = \arg\max_r\{ ||\boldsymbol{\tau}_{ru}|| : r \in \mathcal{B} \}  \label{eqn:optimal_v_distr} \\
& \hspace*{-1.9in} \alpha_{ru}^{(j+1)} = 1/\left(||\boldsymbol{\tau}_{ru}^{(j)}||^2 + \epsilon\right) \label{eqn:optimal_alpha_distr}
\end{eqnarray} 
For the optimal transmit beamformer, the procedures are different than in the centralized case. In~\eqref{eqn:optimal_tau_distr}, $\boldsymbol{\tau}_{ru}$ represents the local optimal decision only. Through information exchange, the optimal transmit beamformer $\mathbf{v}_u$ can be updated by selecting the local optimal transmit beamformer with largest power, as in~\eqref{eqn:optimal_v_distr}. 

Given the expressions for the optimal variables, we can now define the iterative algorithm in Algorithm~\ref{alg:resource_allocation_distributed}. It initializes all users to transmit at full power, initializes the weights, and then iteratively updates the optimal auxiliary variables and transmit beamformers. The new set $\mathcal{T}$ represents the set of local transmit variables $\boldsymbol{\tau}_{ru}$.
\begin{algorithm}
	\caption{Resource Allocation Algorithm for Distributed Operation}\label{alg:resource_allocation_distributed}
	\begin{algorithmic}[1]
		\State{Initialize $\mathbf{v}_{u}$ and $\boldsymbol{\tau}_{ru}$ for all users such that $||\mathbf{v}_{u}||^2=||\boldsymbol{\tau}_{ru}||^2=P_T $}\label{step:initialization_1}
		\State{Initialize $\alpha_{ru}=1/P_T$ for all users}\label{step:initialization_2}
		\State{\textbf{repeat}}
		\State{\quad Update $\Gamma$ using~\eqref{eqn:optimal_gamma_distr}}
		\State{\quad Update $\mathcal{Y}$ using~\eqref{eqn:optimal_y_distr}}
		\State{\quad Update $\mathcal{T}$ using~\eqref{eqn:optimal_tau_distr}}
		\State{\quad Update $\mathcal{V}$ using~\eqref{eqn:optimal_v_distr}}
		\State{\quad Update $\boldsymbol{\alpha}$ using~\eqref{eqn:optimal_alpha_distr}}
		\State {${\mathbf{until}}$ convergence }
	\end{algorithmic}
\end{algorithm}

It is worth noting that, theoretically, we will obtain higher SE or data rate under the centralized operation mode than the distributed mode. This is due to the SINR expression for centralized mode is similar to taking the squared norm of sum of vectors whereas for distributed mode is similar to taking the sum of squared norm of vectors. By sacrificing some performance, the distributed mode gains better fronthaul signaling load and scalability in computation load. This trade-off motivates us to formulate a scheme that combines both centralized and distributed to achieve balance between high performance and low overhead: this semi-distributed operation mode is our focus in the next section. However, as we will discuss later, all three algorithms require substantial real-time information exchange.

Before formulating the semi-distributed approach, it is worth noting that the centralized operation in Section~\ref{chapter:centralize_mode} assumes that all APs are connected to a single CPU. This is clearly infeasible in any network of reasonable size. In this regard, a key difference in the semi-distributed formulation below is the requirement that a CPU only controls a limited number of APs in the network. Furthermore, as seen in Fig.~\ref{fig:network_model}, a user may be served by APs controlled by different CPUs. 

\subsection{Resource Allocation for Semi-Distributed Operation}

We consider the network in Section~\ref{subsection:network_model} where multiple CPUs are deployed in the network where each CPU $q$ is connected with disjoint sets of APs $\mathcal{B}_q$. Let $\mathcal{Q}$ denote the set of CPUs. The APs serve as RRHs and forward their received signal to their corresponding CPU. Each CPU carries out tasks such as channel estimation and resource allocation individually. Then they will send the data to a higher-level CPU (or core network, as in Fig.~\ref{fig:network_model}) for data detection. We remark that CPU $q$ and its set of associated APs in set $\mathcal{B}_q$ jointly operate in centralized mode whereas the set of CPUs $\mathcal{Q}$ and the higher-level CPU jointly operate in a distributed mode. In summary, a higher-level CPU distributes the tasks to many CPUs that are, each, in charge of a set of APs.

We define the set of users that can be scheduled by CPU $q$ as $\mathcal{E}_q$. Even though $\mathcal{B}_q$ may be disjoint sets, $\mathcal{E}_q$ can still have overlapping users due to the potential overlap between $\mathcal{E}_r$, the set of users connected with AP $r$. Similarly, let $\mathcal{D}_u$ be the cluster of CPUs that user $u$ can possibly be served by. Since $|\mathcal{C}_u| > 0$, every user has at least one CPU in its cluster. Finally, we define $\mathcal{C}_{qu} = \mathcal{C}_u \cap \mathcal{B}_q$ as the set of APs that belong to CPU $q$ and also to the cluster for user $u$.

For the resource allocation problem and semi-distributed operation mode, the only change from Section~\ref{subsection:resource_alloc_distr} is the indexing and dimension change of the identity matrix from $\mathbf{I}_M$ to $\mathbf{I}_{|\mathcal{C}_{qu}|M}$. The change from index $ru$ to $qu$ denotes the vertical concatenation for all $r \in \mathcal{C}_{qu}$. For instance, the psuedo-metric $\mathrm{SINR}_{qu}$ that represents the local SINR for user $u$ with respect to CPU $q$
\begin{equation}
	\begin{split}
	\mathrm{SINR}_{qu} & = \boldsymbol{\tau}_{qu}^H \mathbf{H}_{qu}^H
	\Bigl( \sigma^2 \mathbf{I}_{|\mathcal{C}_{qu}|M} + \\ & \hspace*{0.2in} \sum_{u' \in \mathcal{U} \backslash u} \mathbf{H}_{qu'}  \mathbf{v}_{u'} \mathbf{v}_{u'}^H \mathbf{H}_{qu'}^H \Bigl)^{-1} 
	\mathbf{H}_{qu}  \boldsymbol{\tau}_{qu}
	\end{split}
\end{equation}
where channel $\mathbf{H}_{qu} \in \mathbb{C}^{M|\mathcal{C}_{qu}| \times N}$ is the vertical concatenation for all channels $\mathbf{H}_{ru}$ where $r \in \mathcal{C}_{qu}$. The local optimal decision $\boldsymbol{\tau}_{qu}$ is defined in the same way as $\boldsymbol{\tau}_{ru}$ except in this case it is local with respect to CPU $q$ instead of AP $r$.

The WSR optimization problem is defined as
\begin{subequations}
	\begin{align}
		\max_{ \mathcal{V}}\quad & \displaystyle\sum\limits_{q \in \mathcal{Q}}\sum\limits_{u \in \mathcal{E}_q}
		\delta_{qu} \log\left( 1 + 
		\gamma_{qu}
		\right) 
		\\
		\text{s.t.}\quad 
		& \gamma_{qu} = \mathrm{SINR}_{qu}
		\\
		&\displaystyle\sum\limits_{u \in \mathcal{E}_q} \alpha_{qu} ||\boldsymbol{\tau}_{qu}||^2 \le |\mathcal{B}_q| M
		\\
		& ||\boldsymbol{\tau}_{qu}||^2 \le P_T
	\end{align}
\end{subequations}

Applying the same techniques of transformation and solving derivatives, we obtain the following formulas for optimal auxiliary variables and transmit beamformer,
\begin{equation}\label{eqn:optimal_gamma_semi_distr}
	\gamma_{qu}^*=\boldsymbol{\tau}_{qu}^H \mathbf{H}_{qu}^H ( \sigma^2\mathbf{I}_{|\mathcal{C}_{qu}|M} + 
	\sum_{u' \in \mathcal{U} \backslash u} \mathbf{H}_{qu'} \mathbf{v}_{u'} \mathbf{v}_{u'}^H \mathbf{H}_{qu'}^H ) ^{-1}
	\mathbf{H}_{qu} \boldsymbol{\tau}_{qu}
\end{equation}
\begin{equation}
	\begin{split}
	\hspace*{-0.5in} \mathbf{y}_{qu}^* & = \sqrt{\delta_{qu}(1+\gamma_{qu})}
	\Bigl( \sigma^2\mathbf{I}_{|\mathcal{C}_{qu}|M} + \mathbf{H}_{qu} \boldsymbol{\tau}_{qu} \boldsymbol{\tau}_{qu}^H \mathbf{H}_{qu}^H \\
	& \hspace*{0.3in}+\sum_{u' \in \mathcal{U} \backslash u} \mathbf{H}_{qu'} \mathbf{v}_{u'} \mathbf{v}_{u'}^H \mathbf{H}_{qu'}^H 
	\Bigl) ^{-1}
	\mathbf{H}_{qu} \boldsymbol{\tau}_{qu} \label{eqn:optimal_y_semi_distr}
	\end{split}
\end{equation}
\begin{equation}
	\begin{split}
	\hspace*{-0.40in} \boldsymbol{\tau}_{qu}^* & = 
	\sqrt{\delta_{qu}(1+\gamma_{qu})} 
	\Bigl( (\lambda_{q}^* \alpha_{qu} + \mu_{qu}^*) \mathbf{I}_N \\
	& \hspace*{0.2in}+ \sum_{q' \in \mathcal{Q}} \sum_{u' \in \mathcal{E}_{q'}} \mathbf{H}_{q'u}^H \mathbf{y}_{q'u'} 
	\mathbf{y}_{q'u'}^H \mathbf{H}_{q'u} \Bigl)^{-1}
	\mathbf{H}_{qu}^H \mathbf{y}_{qu} \label{eqn:optimal_tau_semi_distr}
	\end{split}
\end{equation}
\begin{equation}\label{eqn:optimal_v_semi_distr}
	\hspace*{-0.65in} \mathbf{v}_{u}^* =  \boldsymbol{\tau}_{q^*u} \mathrm{\ where\ } q^*= \arg\max_q \{ ||\boldsymbol{\tau}_{qu}|| : q \in \mathcal{Q} \}
\end{equation}
\begin{equation}\label{eqn:optimal_alpha_semi_distr}
	\hspace*{-1.95in} \alpha_{qu}^{(j+1)} = 1/\left(||\boldsymbol{\tau}_{qu}^{(j)}||^2 + \epsilon\right)
\end{equation}

Finally, the resource allocation algorithm for semi-distributed operation mode is given in Algorithm~\ref{alg:resource_allocation_semi_distributed}. 
\begin{algorithm}
	\caption{Resource Allocation Algorithm for Semi-Distributed Operation Mode}\label{alg:resource_allocation_semi_distributed}
	\begin{algorithmic}[1]
		\State{Initialize $\mathbf{v}_{u}$ and $\boldsymbol{\tau}_{qu}$ for all users such that $||\mathbf{v}_{u}||^2=||\boldsymbol{\tau}_{qu}||^2=P_T $}\label{step:initialization_1}
		\State{Initialize $\alpha_{qu}=1/P_T$ for all users}\label{step:initialization_2}
		\State{\textbf{repeat}}
		\State{\quad Update $\Gamma$ using~\eqref{eqn:optimal_gamma_semi_distr}}
		\State{\quad Update $\mathcal{Y}$ using~\eqref{eqn:optimal_y_semi_distr}}
		\State{\quad Update $\mathcal{T}$ using~\eqref{eqn:optimal_tau_semi_distr}}
		\State{\quad Update $\mathcal{V}$ using~\eqref{eqn:optimal_v_semi_distr}}
		\State{\quad Update $\boldsymbol{\alpha}$ using~\eqref{eqn:optimal_alpha_semi_distr}}
		\State {${\mathbf{until}}$ convergence }
	\end{algorithmic}
\end{algorithm}

Although we have developed resource allocation algorithms for the distributed and semi-distributed operation mode, this solution still has some practical issues. For instance, in order to calculate the complete interference pattern, AP $r$ or CPU $q$ need the channel information not only for users in its cluster but also for all users in the network. The information for those "non-local" channels are impractical to obtain. In addition, the algorithm involves information exchange of transmit decisions between APs or CPUs which cause scalability issues. However, having explored some of the possible options, we are in a position to present a key contribution of this paper: decentralized resource allocation with multiple CPUs.

\section{Decentralized Resource Allocation}\label{chapter:decentralize_mode}

\subsection{Motivation}
In the previous section, we developed resource allocation algorithms for distributed and semi-distributed operation mode under some assumptions. Importantly, the algorithms developed are FP-based which requires information of the whole network to calculate the interference. We note that running the iterative algorithm to make resource allocation decisions a pure computational process and there is no physical signal sent during each iteration. Thus, we cannot obtain the interference pattern during each iteration from received signals; we can only calculate them through global information which is impractical. 

There are two main reasons why this approach is impractical: first, the AP (or CPU depending on the operation mode) needs to know the resource allocation decisions, such as user scheduling and power allocation, made by other APs from the previous iteration. This requires information exchange between the APs and this exchange scales quadratically with number of APs.  Second, to calculate the interference we require global CSI. AP $r$ not only needs to know the channel to users in $\mathcal{E}_r$, but also the channel to users in $\mathcal{B} \backslash \mathcal{E}_r$. The channel estimation for the latter users is impractical because they will experience strong contamination from co-pilot users that are closer to the AP than them. Finally, the computation complexity of calculating the interference scale with number of users in the whole network which has bad scalability when working with dense population. 

To address these issues, we now propose a decentralized implementation of the resource allocation algorithm that only require \textit{local} information instead of global information to address above problems.

\subsection{Problem Analysis}\label{subsection:problem_analysis_decentralize}
Our goal is to eliminate the need for global information and estimate, to the best of our ability, the overall interference pattern with local information only. Local information for an AP $r$ only includes its own resource allocation decisions, such as $\boldsymbol{\tau}_{ru}$, and the CSI to its users in $\mathcal{E}_r$.

We first analyze the case of distributed operation mode. If we separate the local and non-local interference term from the SINR in~\eqref{eqn:sinr_distr}, the metric can be written in a form of 
\begin{equation}
	\begin{split}
	\mathrm{SINR}_{ru} & = \boldsymbol{\tau}_{ru}^H \mathbf{H}_{ru}^H
	\Bigl( \sigma^2 \mathbf{I}_{M} 
	+\sum_{u' \in \mathcal{E}_r \backslash u} \mathbf{H}_{ru'}  \mathbf{v}_{u'} \mathbf{v}_{u'}^H \mathbf{H}_{ru'}^H \\
	& \hspace*{0.3in} +\sum_{u' \in \mathcal{U} \backslash \mathcal{E}_r} \mathbf{H}_{ru'}  \mathbf{v}_{u'} \mathbf{v}_{u'}^H \mathbf{H}_{ru'}^H \Bigl)^{-1} 
	\mathbf{H}_{ru}  \boldsymbol{\tau}_{ru}
	\end{split}
\end{equation}
Since we wish to eliminate information exchange, AP $r$ is only aware of the local decision variable $\boldsymbol{\tau}_{ru}$ but not $\mathbf{v}_{u}$ which is what user $u$ actually uses to transmit. The SINR can be further reformulated to
\begin{equation}
\begin{split}
	\hspace*{-0.10in} \mathrm{SINR}_{ru} = \boldsymbol{\tau}_{ru}^H \mathbf{H}_{ru}^H
	\Bigl( \sigma^2 \mathbf{I}_{M} 
	\\  & \hspace*{-1.20in} +\sum_{u' \in \mathcal{E}_r \backslash u} \mathbf{H}_{ru'}  \boldsymbol{\tau}_{ru'} \boldsymbol{\tau}_{ru'}^H \mathbf{H}_{ru'}^H 
	 +\mathbf{N}_{ru} \Bigl)^{-1} 
	\mathbf{H}_{ru}  \boldsymbol{\tau}_{ru}
\end{split}
\end{equation}
where the summation over set $\mathcal{E}_r \backslash u$ is the local interference term and $\mathbf{N}_{ru}$ represents the an \textit{estimate} of the sum of non-local interference terms. Assuming the pathloss and shadowing for non-local channels are known because they are previously obtained and used for user-centric clustering of cell-free network (as in Section~\ref{subsection:network_model}), we can \textit{statistically approximate} the sum of non-local interference terms as 
\begin{equation}
	\mathbf{N}_{ru} = \sum_{u' \in \mathcal{U} \backslash u} P_T \mathrm{p}_{ru'} \psi_{ru'} \beta(d_{ru'}) \mathbf{I}_M
\end{equation}
where $\psi_{ru} \beta(d_{ru})$ is the large scale channel statistics, $P_T$ is the maximum transmit power, and $\mathrm{p}_{ru'}$ is the probability that user $u'$ is scheduled in the network by any AP excluding AP $r$. These three terms captures the non-local channel, transmit beamformer, and scheduling variable terms respectively. Recall that even if user $u'$ belongs to $\mathcal{E}_r$, it can still contribute as interference if it is not scheduled on AP $r$ but on some other APs. For the probabilistic scheduling variable, there are many schemes we can consider such as uniform probability. In our case we will use
\begin{equation}
	\mathrm{p}_{ru'} = \sum_{r' \in \mathcal{C}_{u'} \backslash r} \frac{M}{|\mathcal{E}_{r'}|}
\end{equation}
The non-local interference approximation $\mathbf{N}_{ru}$ is a constant diagonal matrix. In the results section, we will use simulation to scale this value and conclude that, in fact, the performance is not very sensitive to the chosen probability scheme.

For simplicity, we merge $\mathbf{N}_{ru}$ with the diagonal noise term to form the psuedo decentralized SINR metric
\begin{equation}\label{eqn:sinr_local_fp}
	\begin{split}
	\mathrm{SINR}_{ru}^{\mathrm{psuedo}} & = \boldsymbol{\tau}_{ru}^H \mathbf{H}_{ru}^H
	\Bigl( \Tilde{\sigma}^2_{ru} \mathbf{I}_{M}  \\
	& +\sum_{u' \in \mathcal{E}_r \backslash u} \mathbf{H}_{ru'}  \boldsymbol{\tau}_{ru'} \boldsymbol{\tau}_{ru'}^H \mathbf{H}_{ru'}^H \Bigl)^{-1} 
	\mathbf{H}_{ru}  \boldsymbol{\tau}_{ru}
	\end{split}
\end{equation}
where $\Tilde{\sigma}^2_{ru}$ represents the combined noise and non-local interference. Essentially~\eqref{eqn:sinr_local_fp} creates a psuedo-SINR metric that involves only the local interference and the relative power of the non-local interference. We do not claim that our approximation accurately captures the interference pattern but it serves as a compensation to the amount of interference that we did not consider in the calculation. We only use the psuedo metric as a proxy for our resource allocation algorithm because we want the solution to be decentralized. In the end, the system performance is still evaluated with real SINR based rates.

Using~\eqref{eqn:sinr_local_fp}, we construct a WSR objective function and formulate the following resource allocation problem to be solved at each AP. At AP $r$, we have
\begin{subequations}\label{eqn:resource_alloc_problem_distr_local_fp}
	\begin{align}
		\max_{ \mathcal{T}_r}\quad & \sum\limits_{u \in \mathcal{E}_r}
		\delta_{ru} \log\left( 1 + 
		\gamma_{ru}
		\right) 
		\\
		\text{s.t.}\quad 
		& \gamma_{ru} = \mathrm{SINR}_{ru}^{\mathrm{psuedo}}
		\\
		&\displaystyle\sum\limits_{u \in \mathcal{E}_r} \alpha_{ru} ||\boldsymbol{\tau}_{ru}||^2 \le M
		\\
		& ||\boldsymbol{\tau}_{ru}||^2 \le P_T
	\end{align}
\end{subequations}
\eqref{eqn:resource_alloc_problem_distr_local_fp} is a decentralized version of~\eqref{eqn:resource_alloc_problem_distr}. The optimal variable set $\mathcal{V}$ no longer needs to be solved jointly among all APs. Instead, $\mathcal{T}_r$ which represents the set of $\boldsymbol{\tau}_{ru}$ for $u \in \mathcal{E}_r$ is solved locally on AP $r$. The psuedo SINR metric decouples the problem so that~\eqref{eqn:resource_alloc_problem_distr} can be split into $|\mathcal{B}|$ subproblems~\eqref{eqn:resource_alloc_problem_distr_local_fp} and each solved individually; thus no information exchange between APs is needed.

\subsection{Resource Allocation for Distributed Operation}\label{subsection:resource_alloc_distr_decentralized}
We employ a FP-based iterative optimization approach to find a optimum for this WSR maximization problem. Since similar procedures have already been analyzed in Section~\ref{subsection:recourse_allocation_cent} and~\ref{subsection:resource_alloc_distr}, we present fewer details. After a series of Lagrange Dual Transform, Quadratic Transform, and partial derivatives, we obtain the following expressions for optimal auxiliary variables and local transmit beamformer:
\begin{equation}\label{eqn:optimal_gamma_distr_local_fp}
	\gamma_{ru}^*=\boldsymbol{\tau}_{ru}^H \mathbf{H}_{ru}^H ( \Tilde{\sigma}^2_{ru} \mathbf{I}_M + 
	\sum_{u' \in \mathcal{E}_r \backslash u} \mathbf{H}_{ru'} \boldsymbol{\tau}_{ru'} \boldsymbol{\tau}_{ru'}^H \mathbf{H}_{ru'}^H ) ^{-1}
	\mathbf{H}_{ru} \boldsymbol{\tau}_{ru}
\end{equation}
\begin{equation}\label{eqn:optimal_y_distr_local_fp}
	\begin{split}
	\hspace*{-0.225in} \mathbf{y}_{ru}^*  =  \sqrt{\delta_{ru}(1+\gamma_{ru})}
	\Bigl( \Tilde{\sigma}^2_{ru} \mathbf{I}_M \\
	& \hspace*{-1.00in} +\sum_{u' \in \mathcal{E}_r} \mathbf{H}_{ru'} \boldsymbol{\tau}_{ru'} \boldsymbol{\tau}_{ru'}^H \mathbf{H}_{ru'}^H 
	\Bigl) ^{-1}
	\mathbf{H}_{ru} \boldsymbol{\tau}_{ru}
	\end{split}
\end{equation}
\begin{equation}\label{eqn:optimal_tau_distr_local_fp}
	\begin{split}
	\hspace*{-0.15in} \boldsymbol{\tau}_{ru}^*  = 
	\sqrt{\delta_{ru}(1+\gamma_{ru})} 
	\Bigl( (\lambda_{r}^* \alpha_{ru} + \mu_{ru}^*) \mathbf{I}_N \\
	& \hspace*{-1.50in} + \sum_{u' \in \mathcal{E}_{r}} \mathbf{H}_{ru'}^H \mathbf{y}_{ru'} 
	\mathbf{y}_{ru'}^H \mathbf{H}_{ru'} \Bigl)^{-1}
	\mathbf{H}_{ru}^H \mathbf{y}_{ru}
	\end{split}
\end{equation}
\begin{equation}\label{eqn:optimal_v_distr_local_fp}
	\hspace*{-0.675in} \mathbf{v}_{u}^* = \boldsymbol{\tau}_{r^*u} \mathrm{\ where\ } r^* = \arg\max_r \{ ||\boldsymbol{\tau}_{ru}|| : r \in \mathcal{B} \}
\end{equation}
\begin{equation}\label{eqn:optimal_alpha_distr_local_fp}
	\hspace*{-1.90in} \alpha_{ru}^{(j+1)} = 1/\left(||\boldsymbol{\tau}_{ru}^{(j)}||^2 + \epsilon\right)
\end{equation}

Given the expressions for the optimal variables, we propose the iterative algorithm in Algorithm~\ref{alg:resource_allocation_distributed_decentralized}. It initializes all users to transmit at full power, initializes the weights, and then iteratively update the optimal auxiliary variables and local transmit beamformers. Unlike Algorithm~\ref{alg:resource_allocation_distributed}, $\mathcal{V}$ is not updated using~\eqref{eqn:optimal_v_distr_local_fp} at every iteration due to the decentralized setup without information exchange. Instead it is updated after the resource allocation algorithm, where we evaluate the overall system performance with real SINR based rates.
\begin{algorithm}
	\caption{Decentralized Resource Allocation Algorithm for Distributed Operation}\label{alg:resource_allocation_distributed_decentralized}
	\begin{algorithmic}[1]
		\State{Initialize $\mathbf{v}_{u}$ and $\boldsymbol{\tau}_{ru}$ for all users such that $||\mathbf{v}_{u}||^2=||\boldsymbol{\tau}_{ru}||^2=P_T $}\label{step:initialization_1}
		\State{Initialize $\alpha_{ru}=1/P_T$ for all users}\label{step:initialization_2}
		\State{\textbf{repeat}}
		\State{\quad Update $\Gamma_r$ using~\eqref{eqn:optimal_gamma_distr_local_fp}} for all $r \in \mathcal{B}$
		\State{\quad Update $\mathcal{Y}_r$ using~\eqref{eqn:optimal_y_distr_local_fp}} for all $r \in \mathcal{B}$
		\State{\quad Update $\mathcal{T}_r$ using~\eqref{eqn:optimal_tau_distr_local_fp}} for all $r \in \mathcal{B}$
		\State{\quad Update $\boldsymbol{\alpha}_r$ using~\eqref{eqn:optimal_alpha_distr_local_fp}} for all $r \in \mathcal{B}$
		\State {${\mathbf{until}}$ convergence }
	\end{algorithmic}
\end{algorithm}

\subsection{Resource Allocation for Semi-Distributed Operation}
In a similar fashion as Section~\ref{subsection:problem_analysis_decentralize} and~\ref{subsection:resource_alloc_distr_decentralized}, we can formulate a decentralized resource allocation algorithm under the semi-distributed operation mode where decisions are made by CPUs, not APs. Now, local information will no longer limited to an AP but to a CPU $q$. The psuedo decentralized SINR metric is defined as
\begin{equation}
	\begin{split}
	\mathrm{SINR}_{qu}^{\mathrm{psuedo}} & = \boldsymbol{\tau}_{qu}^H \mathbf{H}_{qu}^H
	\Bigl( \mathbf{N}_{qu} \\
	& +\sum_{u' \in \mathcal{E}_q \backslash u} \mathbf{H}_{qu'}  \boldsymbol{\tau}_{qu'} \boldsymbol{\tau}_{qu'}^H \mathbf{H}_{qu'}^H \Bigl)^{-1} 
	\mathbf{H}_{qu}  \boldsymbol{\tau}_{qu}
	\end{split}
\end{equation}
where the combined noise and non-local inteference $\mathbf{N}_{qu}$ is defined as
\begin{equation}
	\mathbf{N}_{qu} = \sigma^2_{qu} \mathbf{I}_{|\mathcal{C}_{qu}|M} + \mathrm{diag} \{ \mathbf{N}_{q,ru} : r \in \mathcal{B}_q \}
\end{equation}
\begin{equation}
	\mathbf{N}_{q,ru} = \sum_{u' \in \mathcal{U} \backslash u} P_T \mathrm{p}_{qu'} \psi_{ru'} \beta(d_{ru'}) \mathbf{I}_M
\end{equation}
The term $\mathrm{diag} \{ \mathbf{N}_{q,ru} : r \in \mathcal{B}_q \}$ denotes the block diagonal matrix with the diagonals $\mathbf{N}_{qu}$ for all $r \in \mathcal{B}_q$. For the probabilistic scheduling variable, we will use
\begin{equation}
	\mathrm{p}_{qu'} = \sum_{q' \in \mathcal{D}_{u'} \backslash q} \frac{|\mathcal{B}_q|M}{|\mathcal{E}_{q'}|}
\end{equation}
Again, there are other schemes for defining this value but it is not the focus of this paper. The corresponding resource allocation problem, to be solved at every CPU, is given by
\begin{subequations}    
	\begin{align}
		\max_{ \mathcal{T}_q}\quad & \sum\limits_{u \in \mathcal{E}_q}
		\delta_{qu} \log\left( 1 + 
		\gamma_{qu}
		\right) 
		\\
		\text{s.t.}\quad 
		& \gamma_{qu} = \mathrm{SINR}_{qu}^{\mathrm{psuedo}}
		\\
		&\displaystyle\sum\limits_{u \in \mathcal{E}_q} \alpha_{qu} ||\boldsymbol{\tau}_{qu}||^2 \le |\mathcal{B}_q| M
		\\
		& ||\boldsymbol{\tau}_{qu}||^2 \le P_T
	\end{align}
\end{subequations}
and we use a series of transforms and partial derivatives to obtain the following expressions for optimal auxiliary variables and local transmit beamformer,
\begin{equation}\label{eqn:optimal_gamma_semi_distr_local_fp}
	\gamma_{qu}^*=\boldsymbol{\tau}_{qu}^H \mathbf{H}_{qu}^H ( \mathbf{N}_{qu} + 
	\sum_{u' \in \mathcal{E}_q \backslash u} \mathbf{H}_{qu'} \boldsymbol{\tau}_{qu'} \boldsymbol{\tau}_{qu'}^H \mathbf{H}_{qu'}^H ) ^{-1}
	\mathbf{H}_{qu} \boldsymbol{\tau}_{qu}
\end{equation}
\begin{equation}\label{eqn:optimal_y_semi_distr_local_fp}
\begin{split}
	\hspace*{-0.15in} \mathbf{y}_{qu}^* =  \sqrt{\delta_{qu}(1+\gamma_{qu})}
	( \mathbf{N}_{qu} 
	\\ &\hspace*{-0.7in} +\sum_{u' \in \mathcal{E}_q} \mathbf{H}_{qu'} \boldsymbol{\tau}_{qu'} \boldsymbol{\tau}_{qu'}^H \mathbf{H}_{qu'}^H 
	) ^{-1}
	\mathbf{H}_{qu} \boldsymbol{\tau}_{qu}
 \end{split}
\end{equation}
\begin{equation}\label{eqn:optimal_tau_semi_distr_local_fp}
	\begin{split}
	\hspace*{-0.15in} \boldsymbol{\tau}_{qu}^*  = 
	\sqrt{\delta_{qu}(1+\gamma_{qu})} 
	\Bigl( (\lambda_{q}^* \alpha_{qu} + \mu_{qu}^*) \mathbf{I}_N \\
	& \hspace*{-1.50in} + \sum_{u' \in \mathcal{E}_{q}} \mathbf{H}_{qu'}^H \mathbf{y}_{qu'} 
	\mathbf{y}_{qu'}^H \mathbf{H}_{qu'} \Bigl)^{-1}
	\mathbf{H}_{qu}^H \mathbf{y}_{qu}
	\end{split}
\end{equation}
\begin{equation}\label{eqn:optimal_v_semi_distr_local_fp}
	\hspace*{-0.70in} \mathbf{v}_{u}^* = \boldsymbol{\tau}_{q^*u} \mathrm{\ where\ } q^* = \arg\max \{ ||\boldsymbol{\tau}_{qu}|| : q \in \mathcal{Q} \}
\end{equation}
\begin{equation}\label{eqn:optimal_alpha_semi_distr_local_fp}
	\hspace*{-1.95in} \alpha_{qu}^{(j+1)} = 1/\left(||\boldsymbol{\tau}_{qu}^{(j)}||^2 + \epsilon\right)
\end{equation}

Finally, the decentralized resource allocation algorithm for semi-distributed operation mode is given in Algorithm~\ref{alg:resource_allocation_semi_distributed_decentralized}. 
\begin{algorithm}
	\caption{Decentralized Resource Allocation Algorithm for Semi-Distributed Operation}\label{alg:resource_allocation_semi_distributed_decentralized}
	\begin{algorithmic}[1]
		\State{Initialize $\mathbf{v}_{u}$ and $\boldsymbol{\tau}_{qu}$ for all users such that $||\mathbf{v}_{u}||^2=||\boldsymbol{\tau}_{qu}||^2=P_T $}\label{step:initialization_1}
		\State{Initialize $\alpha_{qu}=1/P_T$ for all users}\label{step:initialization_2}
		\State{\textbf{repeat}}
		\State{\quad Update $\Gamma_q$ using~\eqref{eqn:optimal_gamma_semi_distr_local_fp}} for all $q \in \mathcal{Q}$
		\State{\quad Update $\mathcal{Y}_q$ using~\eqref{eqn:optimal_y_semi_distr_local_fp}} for all $q \in \mathcal{Q}$
		\State{\quad Update $\mathcal{T}_q$ using~\eqref{eqn:optimal_tau_semi_distr_local_fp}} for all $q \in \mathcal{Q}$
		\State{\quad Update $\boldsymbol{\alpha}_q$ using~\eqref{eqn:optimal_alpha_semi_distr_local_fp}} for all $q \in \mathcal{Q}$
		\State {${\mathbf{until}}$ convergence }
	\end{algorithmic}
\end{algorithm}

\section{System Comparisons} \label{section:system_comparisons}

\subsection{Comparison of Operation}

In this section, we will use abbreviations that is exclusive for this section. The naming is arbitrary and the purpose is for easier repetitive references.

Overall, we proposed four layers in the network: a high-level CPU (Layer 1), multiple low-level CPUs (Layer 1.5), APs/RRHs (Layer 2), and users (Layer 3). Our centralized and distributed networks comprise Layer 1, 2, and 3 whereas Layer 1.5 is unique to our semi-distributed network. Layer 2 uses APs for distributed operation and RRHs for other cases. The links between between Layer 2 and 3 are wireless and all other links are wired.

There are also two stages we consider: an algorithmic stage that determines resource allocation decisions through FP (Stage A) and a physical stage including signal transmission and processing (Stage B). As a side note, the allocation decisions, i.e., scheduling and beamforming, from Stage A are delivered to the users through control channels~\cite{chiang2008}.

For the five modes we proposed, their operation schemes are listed as follow:
\begin{itemize}
        \item Centralized (C): Layer 1 performs Stage A since it is the only layer with a processing unit. In Stage B, Layer 3 transmits the signal to Layer 2. Layer 2 does not process the signal and passes on the signal to Layer 1. The signal is jointly processed and the data is estimated at Layer 1. Despite the combining vector being computed in Layer 1, the receive beamforming is done at Layer 2 because they are equipped with antennas and RF chains.
        \item Distributed (D): Layer 2 performs Stage A since they are equipped with processing units in the distributed operation mode. Information exchange is needed within Layer 2 or aided by Layer 1 because multiple processors are running the algorithm simultaneously and they are unaware of the information at other processors. In Stage B, Layer 3 transmits the signal to Layer 2. Layer 2 computes the receiver vector and estimates the data. The estimated data is then sent to Layer 1 for weighted combining as in {\eqref{eq:signal_combined_distributed}}. Note that the received signal is primarily processed on Layer 2 where they first received. The second stage of processing on Layer 1 ensures coherence amongst the APs and is not our focus.
        \item Semi-Distributed (SD): Layer 1.5 performs Stage A and information exchange as needed within Layer 1.5 or aided by Layer 1. In Stage B, Layer 3 transmits the signal to Layer 2. Layer 2 does not process the signal and passes on the signal to Layer 1.5 where data is estimated. The combining vector is computed on Layer 1.5 but the receiving is still done in Layer 2. The estimated data is then sent to Layer 1 for final weighted combining.
        \item Distributed Decentralized (D-D): Layer 2 performs Stage A without any information exchange. The data transmission and detection process is same as the distributed case.
        \item Semi-Distributed Decentralized (SD-D): Layer 1.5 performs Stage A without any information exchange. The data transmission and detection process is same as the semi-distributed case.
\end{itemize}

The operations of different modes are summarized in Table~\ref{tab:table_of_operation}. The table inputs are numbers which indicate which layer does each tasks. Though channel estimation is not the focus for this paper, it is still included for completeness. There is no loss in optimally on which layer performs channel estimation due to channel independence~\cite{Demir2021}.
	
The main computation and data processing tasks for C, D, and SD schemes are assigned to Layer 1, 2, and 1.5 respectively. We split the tasks into different portions and distribute them to different processors. At one extreme, the task is not split and sent to one processor which would represent C. At the other extreme, the tasks are split into smallest possible portions and sent to all processors which would represent D. SD would lie in between. By introducing the low-level CPUs, we allow disjoint groups of APs to jointly process the signal resulting in better rates.

\begin{table}
\centering
\caption{Table of Operation Comparisons}
\begin{tabular}{ | c | c | c | c | }
    \hline
     & C & D & SD \\ \hline            
    Channel Estimation & 1 & 2 & 1.5 \\ \hline      
    FP-Based Resource Allocation & 1 & 2 & 1.5 \\ \hline  
    Receiver Computation  & 1 & 2 & 1.5 \\ \hline      
    Transmit & 3 & 3 & 3 \\ \hline     
    Receive & 2 & 2 & 2 \\ \hline 
    Local Combine (1st) / Data Estimation & 1 & 2 & 1.5 \\ \hline  
    Combine (2nd) of Data Estimates & - & 1 & 1 \\ \hline  
\end{tabular}\label{tab:table_of_operation}
\end{table}

Another advantage for SD network is convenient implementation and adaptation. For any 3-layer network operating in C or D modes, it can always be converted to a SD network by deploying the low-level CPUs, or Layer 1.5, without any physical reconstruction of other layers. The new 4-layer network enables the SD mode, but does not forbid C or D modes. It can operate on all three modes interchangeably depending on the service requirement. For example, when few users need to be served, it can operate in C mode which provides higher data rate. When more users need to be served, it can operate in SD mode which does not provide as high data rates as C but it incurs less overhead because computation and processing tasks are distributed to the low-level CPUs. Finally, the overhead may even be too much for the CPUs to handle for a denser population of users. The network can operate in D so that the tasks are split and further distributed down to the processors located at the APs.

\subsection{Comparison of Scalability}
Scalability is a crucial aspect of our proposed system is. We continue to use the abbreviations from the previous section.

We first consider fronthaul signaling load, i.e., the traffic on each wired fronthaul link. For mode C, the received signal, comprising $M|\mathcal{B}|$ complex numbers, is sent from Layer 2 to 1 and the receive beamforming vector comprising $M|\mathcal{B}|$ complex numbers is sent from Layer 1 to 2. There are a total of $|\mathcal{B}|$ links between Layer 1 and 2 or $2M$ complex scalars per link (and per time slot). For D, the only difference is that only $|\mathcal{B}|$ complex numbers are sent from Layer 2 to 1 because APs send the estimated data to the CPU instead of the signal vector. This equates to a load of $\frac{M+1}{2}$ complex scalars which is on the same order of magnitude compared to  C.

There is an extra Layer 1.5 for SD so there are more complex numbers to be sent. But at the same time, there are also more links available so the average fronthaul signalling load per link would be in between C and D with exact load depends on design. $M$, the number of antennas at each AP, it expected to be small. Therefore, fronthaul signalling load does not create a scalability issue. However, the main issue for the fronthaul is quantization distortion as mentioned earlier.

As network demand increases, the number of users $|\mathcal{U}|$ and APs $|\mathcal{B}|$ grow. We will evaluate the complexity based on how many complex multiplications are required. One of the main scalability issues for C is complexity of computing the receiver vector with {\eqref{eq:MMSEreceiver_centralized}}. For each receiver vector, an $M|\mathcal{C}_u| \times |\mathcal{U}|$ matrix is multiplied to its Hermitian. Then the inverse is taken on the resulting $M|\mathcal{C}_u| \times M|\mathcal{C}_u|$ matrix. Lastly, the inverted matrix is multiplied with a vector of length $M|\mathcal{C}_u|$. The total number of complex multiplications required by these operations can be obtained by using the framework described in~\cite{Björnson2017}. In short, the matrix multiplication requires $\frac{M^2|\mathcal{C}_u|^2 + M|\mathcal{C}_u|}{2} |\mathcal{U}|$ complex multiplications by utilizing the Hermitian symmetry. The inverse requires $\frac{M^3|\mathcal{C}_u|^3 - M|\mathcal{C}_u|}{3}$ through LDL Cholesky decomposition and the final multiplication by a vector costs $M^2|\mathcal{C}_u|^2$. A simplification can be made is replacing $|\mathcal{U}|$ with $M|\mathcal{B}|$ because non-scheduled users does not contribute in the first matrix multiplication.

For D, the same calculation procedures are followed for each receiver vector. The difference is the dimensions of matrices are reduced to $M$. A user may be scheduled on multiple APs where each needs a different receiver. The number of required receivers to compute is at max $|\mathcal{C}_u|$ because that is the maximum number of APs user $u$ can be scheduled on. For SD, the dimension of matrices are reduced to $M|\mathcal{C}_{qu}|$ depending on which CPU $q$ that user $u$ belong to. Summing over all CPUs belong to $\mathcal{D}_u$ would give the upper bound because user $u$ may not be scheduled on all CPUs in its cluster.

The complexity of computing the complex receiver vector is summarized in Table {\ref{tab:table_of_scalability}}. Some simplifications are made such as removing the non-dominant terms. D has better scalability than C mainly because it uses smaller matrices.

The complexity for SD is not straightforward. Note that $\sum_{q\in \mathcal{D}_u}|\mathcal{C}_{qu}| = |\mathcal{C}_{u}|$ because every AP can be only associated to one low-level CPU in SD. The receiver computation complexity in SD equals to the complexity in C only if $|\mathcal{C}_{qu}|=|\mathcal{C}_{u}|$. Under this scenario, everything would be the same in C and user would benefit from high data rate. By design, a low-level CPU will not be in charge of too many APs or else it would be indifferent from a high-level CPU and there is no point of semi-distribution. As a result, the receiver computation complexity in SD will always be less than the complexity in C.

\begin{table}
\centering
\caption{Table of Overhead Comparisons}
\begin{tabular}{ | c | c | c | }
    \hline
     & Computation Complexity & Information Exchange \\ \hline            
    C & $M^3|\mathcal{C}_u|^2|\mathcal{B}|+M^3|\mathcal{C}_u|^3$ & - \\ \hline      
    D & $M^3|\mathcal{C}_u||\mathcal{B}|+M^3|\mathcal{C}_u|$ & $ \sum\limits_{r\in \mathcal{B}} \sum\limits_{r'\in \mathcal{B}\backslash r} MN|\mathcal{E}_{r}| $ \\ 
     &   &  $+N_{iter} (M+N) |\mathcal{E}_{r'}|$  \\ \hline      
    SD & $ \sum\limits_{q\in \mathcal{D}_u} M^3|\mathcal{C}_{qu}|^2|\mathcal{B}|$ & $ \sum\limits_{q\in \mathcal{Q}} \sum\limits_{r'\in \mathcal{B}\backslash \mathcal{B}_q} MN|\mathcal{E}_{q}| $ \\ 
     & \qquad $+M^3|\mathcal{C}_{qu}|^3$ & $+N_{iter} (M+N) |\mathcal{E}_{r'}|$ \\ \hline
\end{tabular}\label{tab:table_of_scalability}
\end{table}

Despite the better scalability through distributing the computation tasks, the need for information arises during resource allocation (since the processors are spatially distributed and unaware of what others are doing). If we were to use the FP-based algorithm we proposed, each AP needs to acquire the knowledge of auxiliary variables computed on the other APs for every iteration. Table \ref{tab:table_of_scalability} summarizes the total amount of information exchange in terms of number of complex numbers needed for exchange. For D, $\mathbf{v}_{u'}$ and $\mathbf{y}_{r'u'}$ are not known by AP $r$ and they are exchanged for every iteration. The non-local channels $\mathbf{H}_{r'u}$ are also unknown by AP $r$, but it only need to be shared once instead of for every iteration. For SD, it is similar but require fewer exchanges because the CPUs know the information for the APs that they are in charge of. The amount of complex channels need for information exchange becomes an issue as more antennas are deployed into the network. Note that these complex variables are exchanged for every iteration of the algorithm. This incurs an additional overhead for synchronization since the next iteration cannot be run without information from the previous iteration. As a result, we proposed the decentralized versions, D-D and SD-D, that completely eliminate the need for these overheads.

The complexity for the resource allocation algorithm has already being studied in~\cite{ammar2022,shen2018}. It scales with respect to the amount of users that a processor needs to make scheduling decisions for. For example, the high-level CPU needs to make decisions for all users in C, so it scales with $|\mathcal{U}|$. For D, each AP needs to make decisions for users that it is associated to so it scales with $|\mathcal{E}_r|$. Similarly, the algorithm scales with $|\mathcal{E}_q|$ for SD. With the task distributed to more processors, the per-processor algorithmic complexity is decreased.

\section{Numerical Results} \label{chapter:results}

In this section, we present the results of simulations to illustrate the efficacy of the proposed algorithms. We consider a wrap-around structure with 7 hexagonal \textit{virtual} cells. We emphasize that the cellular structure is only to generate of users and APs covering a two dimensional space with no other physical meanings. The APs in a cell region are served by a single CPU. Each virtual cell has radius of 500 m. Users and APs are uniformly distributed with a 20 m exclusive region around the APs. 

We use the COST231 Walfisch-Ikegami model~\cite{Walfisch1988} to define the path loss component at the $f = 1800$ MHz band as $\beta(d_{ru}) = -112.4271 - 38 \mathrm{log}_{10} (d_{ru})$, where $d_{ru}$ is measured in km, and a 4 dB lognormal shadowing. We average our results using Monte Carlo simulations over 500 network topologies. We consider user fairness by 100 time slots (TSs) and averaging over the network performance where we call this long term results. We use proportional fair weights defined as
\begin{equation}\label{eqn:fair_weight}
	\delta_{u}^{(t)}=\frac{1}{\Bar{R}^{(t)}}, \Bar{R}^{(t+1)} = \eta R^{(t)} + (1-\eta) \Bar{R}^{(t)}
\end{equation}
where $\delta_{u}^{(t)}$ is $\delta_{u}$ in TS $t$, $R^{(t)}$ is SE in TS $t$, and $\Bar{R}^{(t)}$ is the long-term SE averaged over previous TSs. We use with a forgetting factor of $\eta=0.2$ which, as measured in Jain's Fairness Index~\cite{Jain1998}, provides fairness of greater than 70 \% among 30 TSs. Note that when we evaluate and compare the system performance, we consider the unweighted sum SE with all the weights equal to one. In essence, the weights only ensure different users get a chance to be scheduled over long term and do not contribute to performance evaluation.

\begin{center}
	\begin{table}[t]
		\centering
		\caption{Table of Simulation Parameters} 
		\begin{tabular}[t]{lcc}
			\hline
			Antennas at AP $(M)$ & 8\\
			Antennas at User $(N)$ & 1\\
			Maximum Transmit Power $(P_T)$ & 23 dBm\\
			Noise Spectral Density & -174 dBm/Hz\\
			Noise Figure & 8 dBm\\
			Noise Bandwidth & 20 MHz \\
			Forgetting Factor $(\eta)$ & 0.2\\
			User-centric Cluster Boundary $(\rho)$ & 0.4 km\\
			Compressive Sensing Parameter $(\epsilon)$ & $\frac{M}{0.9P_T}$\\
		
			\hline
			\label{tab:sim_param}
		\end{tabular}
	\end{table}%
\end{center}

For the simulation, we consider single-antenna users for the following reasons: after a fixed value of $N|\mathcal{U}|$ user antennas in the network, the highest SE is achieved by as if having $N|\mathcal{U}|$ single-antenna users~\cite{li2016}. Spatial multiplexing is not beneficial for our considered system with lots of users. In addition, the industry, such as in 3GPP has shown interest in UE complexity reduction~\cite{3gpp2020}. One of the main focus is to reduce the number of UE antennas. Thus, we consider single-stream single-antenna users in our simulation.

Table~\ref{tab:sim_param} lists the parameters used unless stated otherwise. As mentioned, to focus on the algorithm performance, we assume perfect CSI. We begin with evaluating the performance for the centralized operation mode, using Algorithm~\ref{alg:resource_allocation_centralized}, as it will serve as a baseline for comparisons. 

\begin{figure}
	\centering
	\includegraphics[width=3.5in]{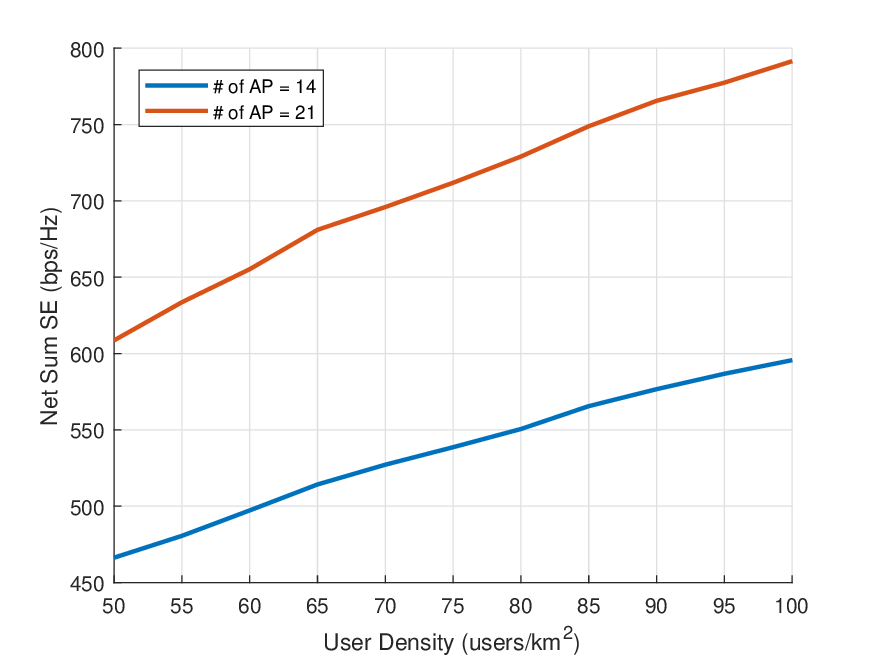}
	\caption{ Sum SE as a function of user density for different number of APs in the network }
	\label{fig:vary_user_cent}
\end{figure}

\begin{figure}[t]
	\centering
	\includegraphics[width=3.5in]{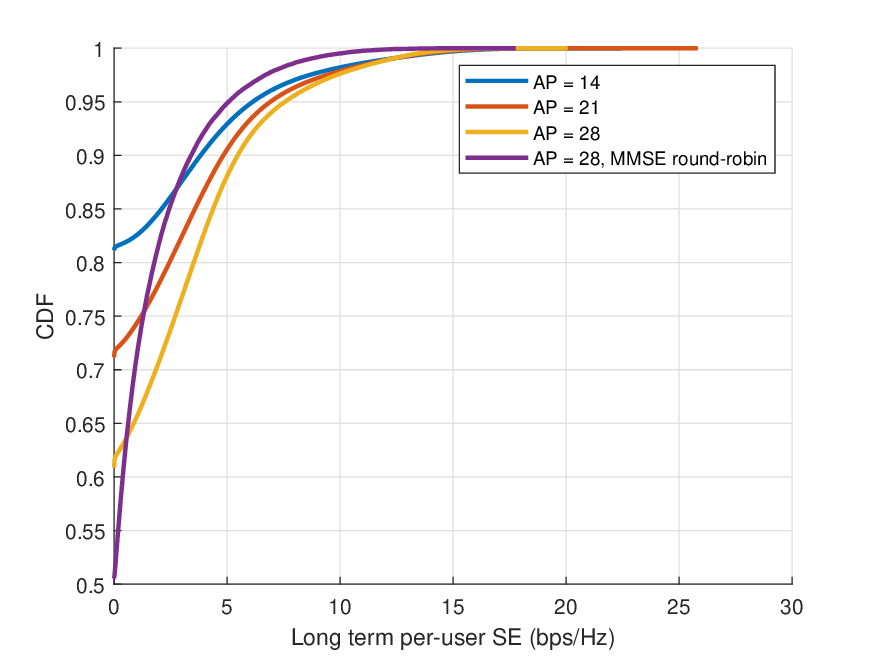}
	\caption{ CDF for Long Term per-user per-TS SE under Centralized Operation Mode} 
	\label{fig:cdf_cent}
\end{figure}

In Fig~\ref{fig:vary_user_cent}, we plot the sum short-term SE (for a single TS) as a function of user density. We observe that as we increase the number of APs in the system, the sum SE increases. This result is expected because as we increase number of APs, we also increase the spatial resources resulting more users getting scheduled achieving higher sum SE. We also notice sum SE increase as users are more densely populated. This is because we are simulating for a single TS only and as we increase user density, there will be more users that are close to the APs resulting a higher SE. We would expect this effect to saturate for large user densities. Next, in Fig~\ref{fig:cdf_cent}, we extend the results of Fig.~\ref{fig:vary_user_cent}, to the long-term rates simulated over 100 TSs. We consider a user density of 100 $\mathrm{users/km}^2$ (448 users) and 14, 21, 28 APs in the network. We plot the CDF of the per-user SE with users unscheduled in a particular TS counted as zero SE. Observing the complementary of CDF, a higher percentage of the users obtain better SE when more APs are deployed. This is due to more users are scheduled with the addition of resources as well as some users are served by more antennas with the addition of APs in their clusters. On the other hand, the drawback of adding APs is the increased interference as more users are scheduled in the network. This explains why the percentage gap closes off for high-SE users. 

For the case of 28 APs, we also simulated for the case of using round-robin scheduling and an MMSE receiver. With 448 users and 224 antennas, the users are uniformly split into 2 groups and they are scheduled on alternating TSs (correspond to 0.5 $y$-intercept). Despite using MMSE receivers, round robin achieves worse data rates which shows the necessity of user scheduling. Another observation can be made is that the $y$-intercept for our proposed method (for AP = 28) is greater than 0.5. This means that not all antennas are used for user multiplexing. Some antennas are used for interference cancellation which improves system performance.

\begin{figure}[t]
	\centering
	\includegraphics[width=3.5in]{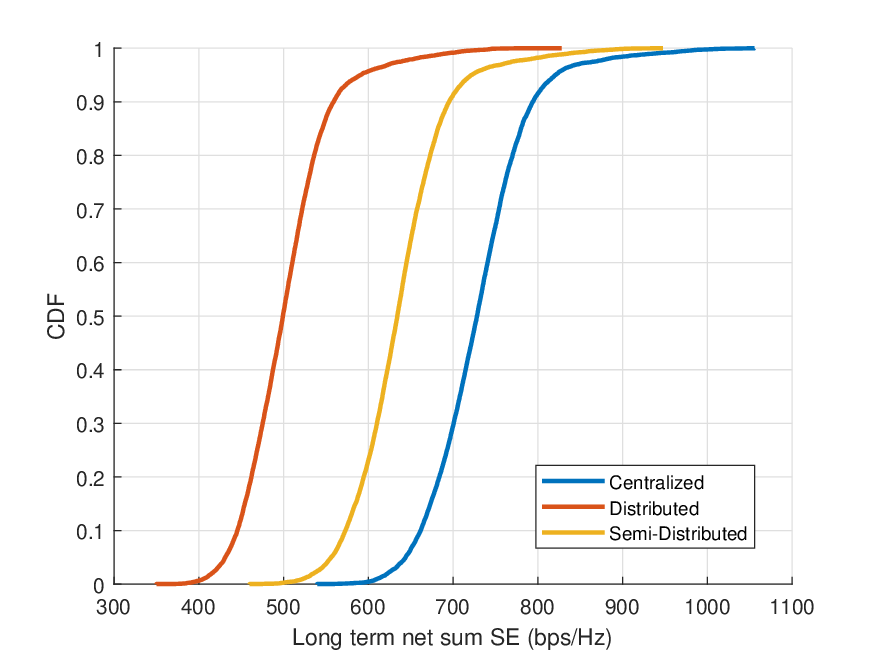}
	\caption{ Comparison of CDF for Long Term Net Sum SE under Different Operation Modes }
	\label{fig:cdf_distributed_compare}
\end{figure}

We use the centralized case to compare the other proposed approaches. For distributed and semi-distributed operation modes, we use Algorithm~\ref{alg:resource_allocation_distributed} and Algorithm~\ref{alg:resource_allocation_semi_distributed} with the same parameters. We consider a user density of 100 $\mathrm{users/km}^2$ (448 users), 28 APs; in Fig~\ref{fig:cdf_distributed_compare}, we compare long-term SE for the three operation modes. The results agree with our theories that the sum SE performance ranking from best to worst is centralized, semi-distributed, and distributed mode. However, as mentioned, the centralized mode sacrifices practicalities for performance whereas distributed mode does the opposite. The semi-distributed lies in between provides a balance between performance and practicalities.

\begin{figure}[t]
	\centering
	\includegraphics[width=7cm]{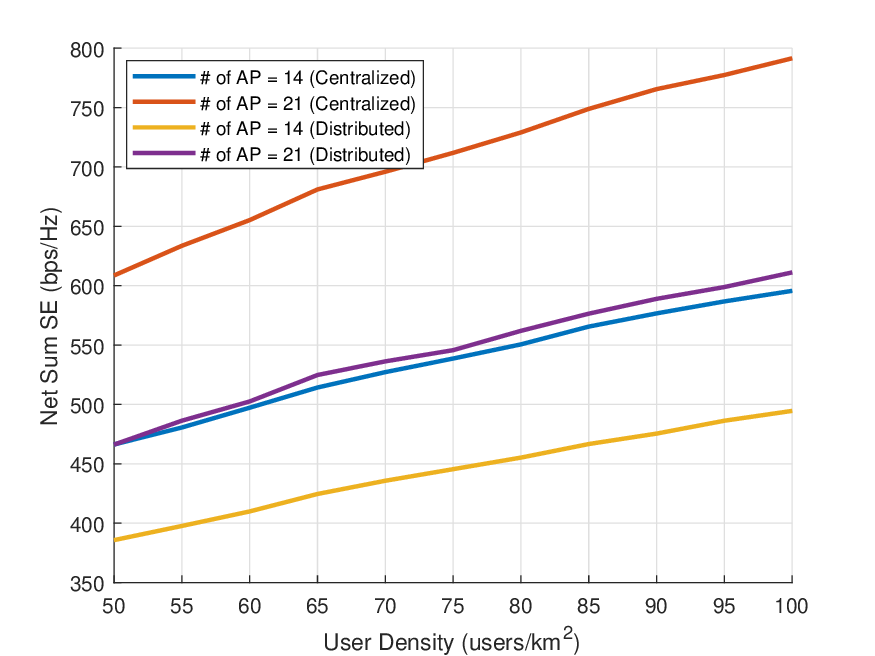}
	\includegraphics[width=7cm]{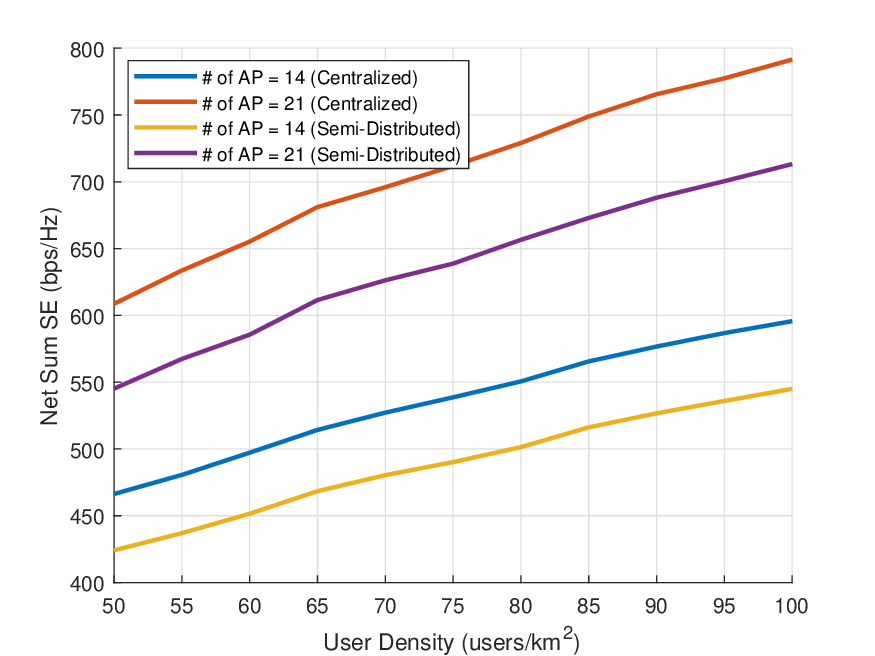}
	\caption{ Sum SE v/s User Density for a single TS for Different Operation Modes }
	\label{fig:vary_user_distributed_compare}
\end{figure}

The results in Fig.~\ref{fig:cdf_distributed_compare} are confirmed by the short-term results shown in Fig~\ref{fig:vary_user_distributed_compare}; here, we plot the sum SE over user density for a single TS. The distributed mode suffers an approximately 17 \% SE decrease for the case of 14 APs and a 23 \% SE decrease for the case of 21 APs. The increasing loss in relative performance with increasing AP density can be explained as follows: recall we made the mathematical intuition that we treat this problem like taking the squared norm. If we take the squared norm over a large number the result would be larger than splitting the large number into many small numbers then taking the sum of squared norms of small numbers. The prior represents the centralized mode and the latter the distributed mode. As a result, the more CPU is distributing the signal processing task, the more penalty in system performance we get. 

For the semi-distributed mode, optimizing AP placement and clustering in relation the to CPU is outside the scope of this paper. Here, we simulate for 7 CPUs corresponding to 7 virtual cells and each has same number of APs in its cluster. From the short term results in Fig~\ref{fig:vary_user_distributed_compare}, it suffers around 10 \% SE decrease for the both cases of 14 APs and 21 APs. Immediately we can tell a less performance penalty compared to distributed mode. The relative loss in SE does not change by much when number of APs in the network changes because the number of total CPUs in the network remains unchanged. This suggests another advantage of using the semi-distributed mode is robust to changes in AP density. But we should be aware this is only true under the case where all CPUs manage the same amount of APs, this conclusion does not carry over to other AP cluster scenarios.

We now consider the decentralized version of resource allocation algorithm under both distributed (Algorithm~\ref{alg:resource_allocation_distributed_decentralized}) and semi-distributed (Algorithm~\ref{alg:resource_allocation_semi_distributed_decentralized}) operation mode are simulated and the results are compared with the results of the operating modes shown previously. We emphasize again that the decentralized psuedo SINR metric is used in the algorithm only; performance evaluation uses the true SINR provided in Section~\ref{subsection:centralized_model} and~\ref{subsection:distributed_model}. 

\begin{figure}[t]
	\centering
	\includegraphics[width=3.5in]{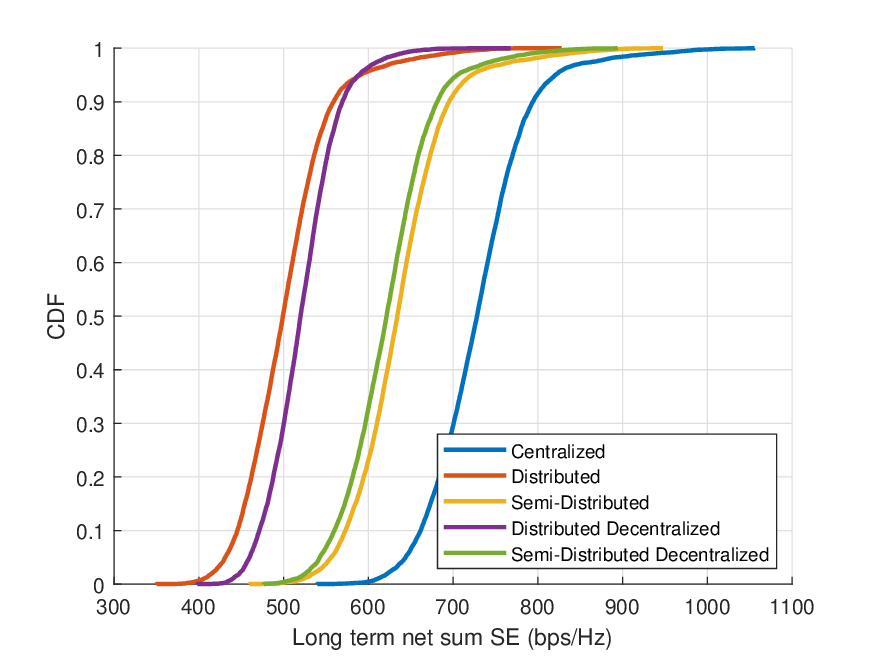}
	\caption{ Comparison of CDF for Long Term Net Sum SE under Different Operation Modes }
	\label{fig:cdf_decentralized_compare}
\end{figure}

In Fig~\ref{fig:cdf_decentralized_compare}, we consider a user density of 100 $\mathrm{users/km}^2$ (448 users), 4 APs/CPU and we simulated for long-term SE with five different operation modes. The CDF results show a decrease in sum SE when using the decentralized algorithm for both distributed and semi-distributed mode. Since the psuedo metric considers local information and static non-local information when computing the interference, it expected that the results obtained with psuedo metric are not the optimum when we evaluate based on true SINR metric. Further, we note that since using the psuedo metric helps us decouple the problem we solve multiple subproblems individually. In other words, the decentralized algorithm allow each APs to make resource allocation decision in a selfish way because they do not have accurate global knowledge. Essentially, we are further sacrificing performance in exchange for better scalability and eliminate the need for information exchange between APs. However, as is clear from the figure, the additional loss in performance is, in fact, minimal. 

From the CDF, we notice that the decentralized algorithm appears to give higher rate over the long term for the distributed operation mode. This is not because the decentralized algorithm makes better resource allocation decisions. It is due to the fact that the weights between the same TS over two scenarios are different. Given the proportional fairness weight {\eqref{eqn:fair_weight}} is inversely proportional to the rate, the users with strong channel gain will be scheduled more frequently in the decentralized case because their weights are penalized less after each TS. This makes an unfair performance comparison and is why we will focus quantitative comparisons for the short term where all weights equal to one.

\begin{figure}[t]
	\centering
	\includegraphics[width=7cm]{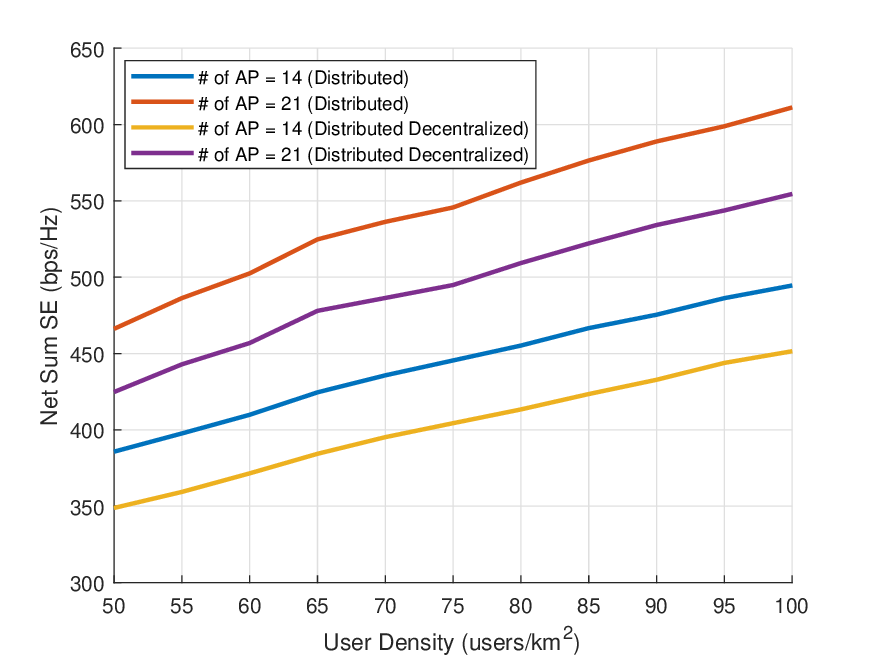}
	\includegraphics[width=7cm]{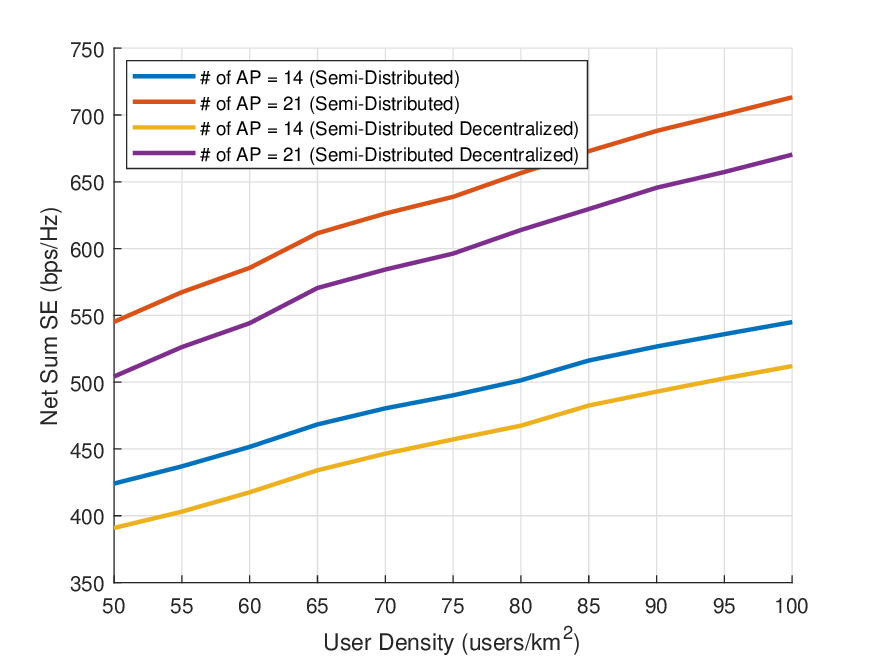}
	\caption{ User Density vs. 1TS Sum SE for Different Operation Modes }
	\label{fig:vary_user_decentralized_compare}
\end{figure}

In Fig~\ref{fig:vary_user_decentralized_compare}, we plot the sum SE over user density and compare the results between using and not using decentralization for both modes. For distributed mode, it suffers an approximately 9 \% SE decrease for the case of 14 APs (2 APs/CPU) and 21 APs (3 APs/CPU). The corresponding number for the semi-distributed mode is a 7 \% SE decrease for the both cases of 14 APs and 21 APs. We notice the performance lost does not vary much when we change the number of APs in the network suggesting that under the decentralization steps is robust under change of AP densities. 

We also observe that semi-distributed mode is penalized less in performance than distributed mode when using the decentralized resource allocation algorithm. This is because there is an embedded information exchange between APs that are connected to the same CPU under semi-distributed operation compared distributed case which has none. Thus, with more known information, the algorithm makes less greedy decisions improving the SE performance. Note, however, that the decisions made by a CPU is still greedy with respect to other CPUs, though not for the APs within its control.

\begin{figure}[t]
	\centering
	\includegraphics[width=3.5in]{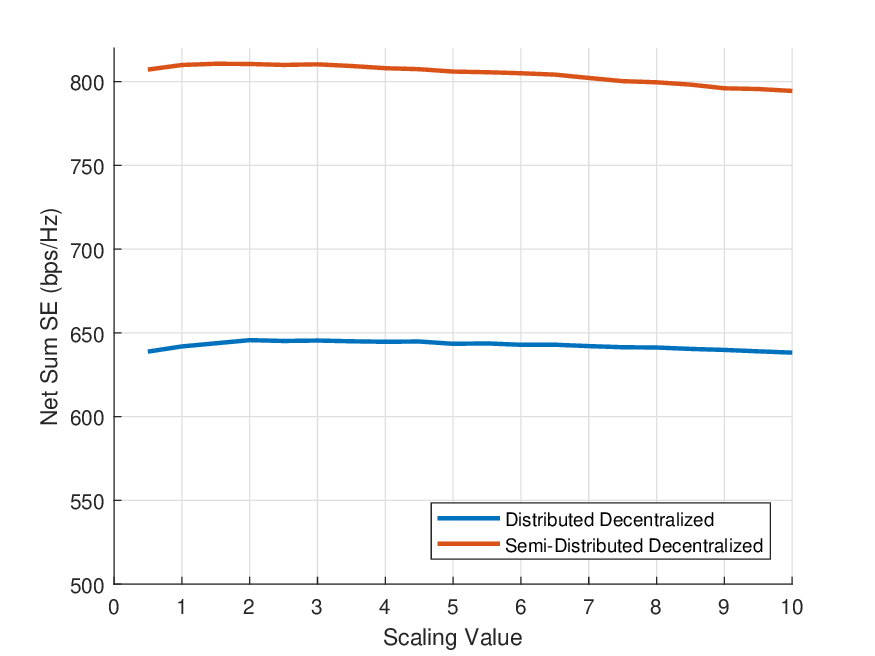}
	\caption{ Scale of Non-local Interference Approximation vs. 1TS Sum SE for Two Different Operation Modes }
	\label{fig:scale_decentralized}
\end{figure}

Finally, we evaluate the impact of scaling the equivalent noise term in the decentralized case. We use unity (no scaling) as a baseline for comparison. In Fig~\ref{fig:scale_decentralized} we change the scaling value from 0.5 to 10; importantly, we note that the sum SE varies by less than 2 \% over this range. A scaling factor of 2 will achieve slightly higher SE for both operation modes but this improvement is very insignificant. We also notice as the scaling factor falls below 1, the performance starts to decrease which suggest the necessity of adding the non-local interference approximation. Overall, the plot suggests it is not worth investing the effort to find better approximations for the non-local interference. Our current scheme, which uses a constant as the approximation and is static over iterations, is adequate. 

\section{Conclusion} \label{chapter:conclusion}

In this paper, we studied the resource allocation optimization problem for a user-centric cell-free network. Our focus was to find an algorithm that would make scheduling and beamforming decisions that result in good overall data rate performance while also ensuring scalability and practicality. Our key contribution was to consider the resource allocation problem in the uplink, a scenario much more complicated than the previously investigated downlink.

First, we used tools of FP and compressive sensing to develop an algorithm that finds the optimal beamformers and user scheduling decisions under the centralized operation mode. By exploiting the multi-layerness in the user-centric cell-free network, we studied the distributed operation mode and proposed the semi-distributed operation mode which is a mix of both centralized and distributed modes. Finally, we proposed using a pseudo-metric for the resource allocation algorithm to obtain a decentralized solution and avoid information exchanges.

In terms of performance, the centralized mode achieved high data rates, but at the cost of heavy fronthaul signaling load, receiver scalability, and computation complexity. By achieving a lower overall data rate over the network, the distributed mode addresses these practical issues. The semi-distributed operation mode combines the benefits of the two modes, achieving competitive data rates while still being relatively more scalable. Additionally, this operation mode is also robust to changes in the number of APs in the system. However, both resource allocation for the distributed and semi-distributed modes are under the assumption that there are information exchanges between the APs, which causes significant overhead. Using the decentralized algorithm resulted in less than 9 \% data rate loss and was robust to changes in AP densities. Moreover, the data rate penalty for the semi-distributed operation mode was less compared to the distributed mode. Hence, we eliminated the need for information exchange between APs while still achieving comparable data rates.

\section*{Acknowledgement\centering}
Prof. Adve would like to acknowledge the support of Ericsson Canada and National Sciences and Engineering Research Council (NSERC) of Canada. The authors appreciates Jingjie Wei, Hussein A. Ammar, and Ahmad Ali Khan for fruitful discussions. The authors also wish to thank the Digital Research Alliance of Canada (alliancecan.ca).

\ifCLASSOPTIONcaptionsoff
  \newpage
\fi

\bibliographystyle{ieeetr}

\begin{thebibliography}{10}

\bibitem{Demir2021}
Özlem Tugfe~Demir, E.~Björnson, and L.~Sanguinetti, ``Foundations of user-centric cell-free massive {MIMO},'' {\em Foundations and Trends{\textregistered} in Signal Processing}, vol.~14, no.~3-4, pp.~162--472, 2021.

\bibitem{khan2020}
A.~A. Khan, R.~S. Adve, and W.~Yu, ``Optimizing downlink resource allocation in multiuser {MIMO} networks via fractional programming and the {H}ungarian algorithm,'' {\em IEEE Trans.~on Wireless Comm.}, vol.~19, no.~8, pp.~5162--5175, 2020.

\bibitem{liu2014}
Y.~Liu, J.~X. Wang, P.~Wang, and R.~K. Yu, ``Joint scheduling and precoding based on {SLNR} criteria in {MU-MIMO} system,'' in {\em Mech.~Comp.~and Cont.~Eng.~III}, vol.~668 of {\em Appl.~Mech.~and Mat.}, Trans Tech Pub.~Ltd., 11 2014.

\bibitem{li2022}
Z.~Li, T.~Gamvrelis, H.~A. Ammar, and R.~Adve, ``Decentralized user scheduling and beamforming in multi-cell {MIMO} networks,'' in {\em ICC 2022 - IEEE International Conference on Communications}, pp.~1980--1985, 2022.

\bibitem{gamvrelis2022}
T.~Gamvrelis, Z.~Li, A.~A. Khan, and R.~S. Adve, ``{SLINR}-based downlink optimization in {MU-MIMO} networks,'' {\em IEEE Access}, vol.~10, pp.~123956--123970, 2022.

\bibitem{shen2018}
K.~Shen and W.~Yu, ``Fractional programming for communication systems—part {II}: Uplink scheduling via matching,'' {\em IEEE Transactions on Signal Processing}, vol.~66, no.~10, pp.~2631--2644, 2018.

\bibitem{ammar2022ucsurvey}
H.~A. Ammar, R.~Adve, S.~Shahbazpanahi, G.~Boudreau, and K.~V. Srinivas, ``User-centric cell-free massive {MIMO} networks: A survey of opportunities, challenges and solutions,'' {\em IEEE Communications Surveys \& Tutorials}, vol.~24, no.~1, pp.~611--652, 2022.

\bibitem{marzetta2016}
T.~L. Marzetta, E.~G. Larsson, H.~Yang, and H.~Q. Ngo, {\em Fundamentals of Massive MIMO}.
\newblock Cambridge University Press, 2016.

\bibitem{ngo2017}
H.~Q. Ngo, A.~Ashikhmin, H.~Yang, E.~G. Larsson, and T.~L. Marzetta, ``Cell-free massive {MIMO} versus small cells,'' {\em IEEE Transactions on Wireless Communications}, vol.~16, no.~3, pp.~1834--1850, 2017.

\bibitem{bjornson2020mmse}
E.~Björnson and L.~Sanguinetti, ``Making cell-free massive {MIMO} competitive with {MMSE} processing and centralized implementation,'' {\em IEEE Transactions on Wireless Communications}, vol.~19, no.~1, pp.~77--90, 2020.

\bibitem{zhang2020b5g}
J.~Zhang, E.~Björnson, M.~Matthaiou, D.~W.~K. Ng, H.~Yang, and D.~J. Love, ``Prospective multiple antenna technologies for beyond 5{G},'' {\em IEEE Journal on Selected Areas in Communications}, vol.~38, no.~8, pp.~1637--1660, 2020.

\bibitem{atzeni2021}
I.~Atzeni, B.~Gouda, and A.~Tölli, ``Distributed precoding design via over-the-air signaling for cell-free massive {MIMO},'' {\em IEEE Transactions on Wireless Communications}, vol.~20, no.~2, pp.~1201--1216, 2021.

\bibitem{kaviani2012}
S.~Kaviani, O.~Simeone, W.~A. Krzymien, and S.~Shamai, ``Linear precoding and equalization for network {MIMO} with partial cooperation,'' {\em IEEE Transactions on Vehicular Technology}, vol.~61, no.~5, pp.~2083--2096, 2012.

\bibitem{masoumi2020}
H.~Masoumi and M.~J. Emadi, ``Performance analysis of cell-free massive {MIMO} system with limited fronthaul capacity and hardware impairments,'' {\em IEEE Transactions on Wireless Communications}, vol.~19, no.~2, pp.~1038--1053, 2020.

\bibitem{Interdonato2020}
G.~Interdonato, M.~Karlsson, E.~Björnson, and E.~G. Larsson, ``Local partial zero-forcing precoding for cell-free massive {MIMO},'' {\em IEEE Transactions on Wireless Communications}, vol.~19, no.~7, pp.~4758--4774, 2020.

\bibitem{dandrea2021}
C.~D'Andrea and E.~G. Larsson, ``User association in scalable cell-free massive {MIMO} systems,'' {\em CoRR}, vol.~abs/2103.05321, 2021.

\bibitem{nguyen2020}
H.~V. Nguyen, V.-D. Nguyen, O.~A. Dobre, S.~K. Sharma, S.~Chatzinotas, B.~Ottersten, and O.-S. Shin, ``On the spectral and energy efficiencies of full-duplex cell-free massive {MIMO},'' {\em IEEE Journal on Selected Areas in Communications}, vol.~38, no.~8, pp.~1698--1718, 2020.

\bibitem{dong2019}
G.~Dong, H.~Zhang, S.~Jin, and D.~Yuan, ``Energy-efficiency-oriented joint user association and power allocation in distributed massive {MIMO} systems,'' {\em IEEE Transactions on Vehicular Technology}, vol.~68, no.~6, pp.~5794--5808, 2019.

\bibitem{lozano2013}
A.~Lozano, R.~W. Heath, and J.~G. Andrews, ``Fundamental limits of cooperation,'' {\em IEEE Transactions on Information Theory}, vol.~59, no.~9, pp.~5213--5226, 2013.

\bibitem{zhu2016}
C.~Zhu and W.~Yu, ``Stochastic analysis of user-centric network {MIMO},'' in {\em 2016 IEEE 17th International Workshop on Signal Processing Advances in Wireless Communications (SPAWC)}, pp.~1--5, 2016.

\bibitem{ammar2022}
H.~A. Ammar, R.~Adve, S.~Shahbazpanahi, G.~Boudreau, and K.~V. Srinivas, ``Downlink resource allocation in multiuser cell-free {MIMO} networks with user-centric clustering,'' {\em IEEE Trans.~on Wireless Comm.}, vol.~21, no.~3, 2022.

\bibitem{Ammar2022distributed}
H.~A. Ammar, R.~Adve, S.~Shahbazpanahi, G.~Boudreau, and K.~V. Srinivas, ``Distributed resource allocation optimization for user-centric cell-free {MIMO} networks,'' {\em IEEE Trans.~on Wireless Comm.}, vol.~21, no.~5, pp.~3099--3115, 2022.

\bibitem{interdonato2019}
G.~Interdonato, P.~Frenger, and E.~G. Larsson, ``Scalability aspects of cell-free massive {MIMO},'' in {\em ICC 2019 - 2019 IEEE International Conference on Communications (ICC)}, pp.~1--6, 2019.

\bibitem{luo2008}
Z.-Q. Luo and S.~Zhang, ``Dynamic spectrum management: Complexity and duality,'' {\em IEEE Journal of Selected Topics in Signal Processing}, vol.~2, no.~1, pp.~57--73, 2008.

\bibitem{candes2007}
E.~Candès, M.~Wakin, and S.~Boyd, ``Enhancing sparsity by reweighted {L1} minimization,'' {\em Journal of Fourier Analysis and Applications}, vol.~14, pp.~877--905, 11 2007.

\bibitem{Björnson2017}
E.~Bj\"{o}rnson, J.~Hoydis, and L.~Sanguinetti, ``Massive {MIMO} networks: {Spectral}, energy, and hardware efficiency,'' {\em Foundations and Trends{\textregistered} in Signal Processing}, vol.~11, no.~3-4, pp.~154--655, 2017.

\bibitem{chiang2008}
M.~Chiang, P.~Hande, T.~Lan, and C.~W. Tan, ``Power control in wireless cellular networks,'' {\em Foundations and Trends® in Networking}, vol.~2, no.~4, pp.~381--533, 2008.

\bibitem{Walfisch1988}
J.~Walfisch and H.~Bertoni, ``A theoretical model of {UHF} propagation in urban environments,'' {\em IEEE Transactions on Antennas and Propagation}, vol.~36, no.~12, pp.~1788--1796, 1988.

\bibitem{Jain1998}
R.~Jain, D.~M. Chiu, and H.~WR, ``A quantitative measure of fairness and discrimination for resource allocation in shared computer systems,'' {\em CoRR}, vol.~cs.NI/9809099, 01 1998.

\bibitem{li2016}
X.~Li, E.~Björnson, S.~Zhou, and J.~Wang, ``Massive {MIMO} with multi-antenna users: When are additional user antennas beneficial?,'' in {\em 2016 23rd International Conference on Telecommunications (ICT)}, pp.~1--6, 2016.

\bibitem{3gpp2020}
"Potential UE complexity reduction features for RedCap," 3GPP, Sophia Antipolis, France, 2020. [Online]. Avaliable: \url{https://www.3gpp.org/ftp/tsg_ran/WG1_RL1/TSGR1_101-e/R1-2003289.zip}.

\end{thebibliography}

\end{document}